\documentclass[useAMS,usenatbib]{mn2e}
\usepackage{graphicx}
\usepackage{times}
\usepackage{amssymb}
\usepackage{amsmath}
\usepackage{multirow}
\usepackage{subfig}
\usepackage{hyphenat}
\usepackage{widetext}
\usepackage{hyperref}
\newcommand{\hompc}{\,h\,{\rm Mpc}^{-1}}
\newcommand{\mpcoh}{\,h^{-1}\,{\rm Mpc}}

\newcommand{\bk}{\boldsymbol{k}}

\newcommand{\bx}{\boldsymbol{x}}
\newcommand{\br}{\boldsymbol{r}}
\newcommand{\bv}{\boldsymbol{v}}

\newcommand{\bC}{\boldsymbol{\mathrm{C}}}

\begin{document}

\title[Peculiar Velocity Forecasts] 
{Cosmological Forecasts for Combined and Next Generation Peculiar Velocity Surveys}

\author[C. Howlett et. al.]{\parbox{\textwidth}{
Cullan Howlett\thanks{Email: cullan.howlett@icrar.org}$^{1,2}$,
Lister Staveley-Smith$^{1,2}$,
Chris Blake$^{2,3}$
}
  \vspace*{4pt} \\ 
$^{1}$International Centre for Radio Astronomy Research, The University of Western Australia, Crawley, WA 6009, Australia. \\
$^{2}$ ARC Centre of Excellence for All-sky Astrophysics (CAASTRO). \\
$^{3}$ Centre for Astrophysics \& Supercomputing, Swinburne University of Technology, P.O.\ Box 218, Hawthorn, VIC \\
}

\pagerange{\pageref{firstpage}--\pageref{lastpage}} \pubyear{2016}
\maketitle
\label{firstpage}

\begin{abstract}
Peculiar velocity surveys present a very promising route to measuring the growth rate of large-scale structure and its scale dependence.
However, individual peculiar velocity surveys suffer from large statistical errors due to the intrinsic scatter in the relations used to infer a galaxy's true distance. In this context we use a Fisher Matrix formalism to investigate the statistical benefits of combining multiple peculiar velocity surveys.
We find that for all cases we consider there is a marked improvement on constraints on the linear growth rate $f\sigma_{8}$. For example, the constraining power of only a few peculiar velocity measurements is such that the addition of the 2MASS Tully-Fisher survey (containing only $\sim2,000$ galaxies) to the full redshift and peculiar velocity samples of the 6-degree Field Galaxy Survey (containing $\sim 110,000$ redshifts and $\sim 9,000$ velocities) can improve growth rate constraints by $\sim20\%$. Furthermore, the combination of the future TAIPAN and WALLABY+WNSHS surveys has the potential to reach a $\sim3\%$ error on $f\sigma_{8}$, which will place tight limits on possible extensions to General Relativity.
We then turn to look at potential systematics in growth rate measurements that can arise due to incorrect calibration of the peculiar velocity zero-point and from scale-dependent spatial and velocity bias. For next generation surveys, we find that neglecting velocity bias in particular has the potential to bias constraints on the growth rate by over $5\sigma$, but that an offset in the zero-point has negligible impact on the velocity power spectrum.
\end{abstract}


\section{Introduction}
The current concordance model of cosmology consists of a universe whose dynamics and geometry can be described using solutions to General Relativity (GR; \citealt{Einstein1916}). In this model, the gravitational evolution of the Universe is caused by an energy-momentum tensor with only four components: radiation, baryonic and dark matter, and dark energy in the form of a cosmological constant. There exists overwhelming support for this consensus cosmological model from observations throughout the expansion history of the universe, including those of the Cosmic Microwave Background (CMB, \citealt{Planck2015a}), supernovae \citep{Freedman2012, Betoule2014}, galaxy lensing \citep{Heymans2012} and the large scale distribution of galaxies \citep{Anderson2014}. However, whilst the inclusion of a cosmological constant recovers the measured expansion rate of the universe and the rate at which structure within it grows, its exact nature remains unknown. Whether dark energy is indeed caused by a cosmological constant or arises instead due to large-scale post-GR modifications to gravity remains one of the fundamental unanswered questions in cosmology.

One of the key observables that allows us to distinguish between different models of gravity and dark energy is the linear growth \textit{rate}, $f=d\,\mathrm{ln}\,D(a)/d\,\mathrm{ln}\,a$, the logarithmic derivative of the linear growth \textit{factor}, $D(a)$, with respect to the scale factor of the universe, $a$. The linear growth factor describes how a density perturbation in the linear regime grows over time. Furthermore, under the assumption of GR, the linear continuity equation can be used to relate the density field, $\delta(\bx,a)$, to the velocity field,$\bv(\bx,a)$, via
\begin{equation}
\nabla \cdot \bv(\bx,a) = -a^{2}H(a)\,d\delta(\bx,a)/{da}.
\end{equation}
$H(a)$ is the cosmology-dependent Hubble parameter. Knowing how a density perturbation evolves with time and using the equation for the linear growth rate, this becomes
\begin{equation}
\nabla \cdot \bv(\bx,a) = -aH(a)f(a)\delta(\bx,a).
\label{eq:cont}
\end{equation}
We can also define the velocity divergence field as, $\theta(\bx,a) = \nabla \cdot \bv(\bx,a)/(aH(a)f(a))$. Comparing this to Eq.~\ref{eq:cont} we can see that, at least on linear scales, the velocity divergence field and the density field are equivalent. However, these two quantities are not physically the same and on non-linear scales they differ.

Hence, measurements of the velocity and density field can be used to place constraints on the linear growth rate. As different cosmological models predict different linear growth rates as a function of scale and time we can also use this to place constraints on the nature of dark energy and gravity. In particular GR predicts a scale-independent linear growth rate that evolves with redshift as $f(z) \approx \Omega_{m}(z)^{\gamma}$, where $\gamma\approx0.55$ \citep{Linder2007}. Measuring a linear growth rate that differs from this prediction, could be seen as a `smoking gun' and be used to advocate alternative theories of gravity.

The tightest current constraints on the growth rate of structure come from measurements of the redshift-space clustering of galaxies. Measured redshifts contain contributions from both the ``Hubble flow" due to the expansion of the universe, and peculiar velocities that arise as galaxies infall towards gravitational potential wells. In terms of the clustering of galaxies this infall creates anisotropies in otherwise isotropic distributions of galaxies when their distances are inferred from the measured redshifts. This effect is called Redshift Space Distortions (RSD, \citealt{Kaiser1987}). Galaxies infalling towards structures perfectly parallel to the line-of-sight will receive maximal additions to their measured redshifts and will appear further away or nearer than they truly are; infall perpendicular to the line-of-sight will result in a measured redshift that is unaffected by the peculiar velocity. The rate of infall is a result of the growth rate of structure and so measurements of the anisotropic clustering, i.e., the multipole moments of the galaxy power spectrum or correlation function (c.f. Eq.~\ref{eq:pkmoments}), can be used to constrain this to high precision. 

Unfortunately, galaxies are biased tracers of the underlying matter distribution and do not trace the gravitational potential wells that source the peculiar velocities exactly. In the linear regime the effect of this ``galaxy bias'' on the galaxy power spectrum or correlation function with respect to the underlying dark matter is completely degenerate with that of RSD and the growth rate of structure. On non-linear scales modelling of the redshift space clustering becomes difficult due to the virial motions of galaxies within their host halos which also contributes to the peculiar velocities. Therefore, although RSD are able to produce tight constraints on the growth rate, their efficacy is limited by how far into the non-linear regime one can model the galaxy clustering; inaccurate modelling on small scales can bias constraints, whilst being too conservative means that the degeneracy with galaxy bias reduces their constraining power.

To overcome the effects of galaxy bias, one alternative is to measure the velocity field directly, as opposed to RSD. From the continuity equation one can see that the velocity field is produced directly by the underlying density perturbation, not by the density field measured using the galaxy sample. Direct measurements of the velocity field can be obtained using empirical measures of a galaxy's true distance compared to the distance inferred from its redshift. The difference between these two measures isolates the galaxy's peculiar velocity. Such empirical relations include the Tully-Fisher \citep{Tully1977} and Fundamental Plane \citep{Dressler1987, Djorgovski1987} relations and the use of supernovae as standard candles \citep{Phillips1993}. These peculiar velocity surveys generally suffer from large statistical errors, due to the intrinsic scatter in the astrophysical relations. High signal to noise requirements also mean that peculiar velocities can only be obtained for low-redshift, local galaxies. However these surveys have seen a resurgence in recent years with larger samples becoming available \citep{Springob2009,Magoulas2012,Hong2014}, and surveys using supernovae remain competitive as the reduced scatter in their luminosity distance relation compensates for the lower number of objects. 

The cosmological benefits of having a peculiar velocity survey rather than a sample containing only redshift measurements can also be expressed as follows. For a survey consisting only of galaxy redshifts, a common technique to recover cosmological information is to convert those redshifts to distances and correlate the locations of pairs of galaxies. This results in a measurement of the power spectrum or correlation function of the density field for that galaxy sample. By way of RSD, this can then be used to measure the growth rate. However, for each galaxy in a peculiar velocity survey we can measure both the peculiar velocity and the redshift/distance of the object. Then, on top of looking at correlations between the locations of pairs of galaxies, we could also correlate their velocities, which gives us a measure of the velocity power spectrum, and cross-correlate the velocity of one galaxy with the position of another, which gives us a measure of the velocity-density cross-power spectrum. This has the benefit of partially cancelling the sample variance between the velocity and density fields and allows us to overcome some of the limitations of cosmic variance.

Peculiar velocities alone, i.e., without the auto- or cross-correlation of the locations of the galaxies in the sample, have already been used to constrain the growth rate \textit{independent} of the galaxy bias. Such measurements have been made using supernovae \citep{Macaulay2012} and the 6dFGSv peculiar velocity sample \citep{Johnson2014}. The benefits of using both the peculiar velocities and redshifts to measure all three potential correlations has been explored theoretically by \cite{Burkey2004} and \cite{Koda2014}, and could be investigated further using current datasets. Both \cite{Burkey2004} and \cite{Koda2014} showed that the addition of a small number of peculiar velocities has the potential to substantially improve growth rate constraints from RSD alone, and allow for a sensitive test of its \textit{scale-dependence}, especially on the largest scales typically outside the regime of redshift-only surveys. They also showed that such constraints will improve with the next generation of peculiar velocity surveys, which push to larger redshifts and cosmological volumes and go deeper in magnitude. 

Beyond this, there now exists sufficient data in overlapping regions of the sky to consider the benefit of combining multiple peculiar velocity surveys, each with their own redshifts and peculiar velocities. In this work we investigate the statistical benefit that arises from combining current and future surveys and explore possible reasons for doing so. This is done by extending the Fisher matrix method used by \cite{Burkey2004} and \cite{Koda2014} to work for multiple measurements of the velocity and density auto- and cross-correlations before forecasting constraints on the growth rate than can be obtained under different scenarios. Our formalism is general enough that it can easily be adapted to accommodate a larger number of overlapping surveys, and observables beyond the density and velocity fields. For example, one could just as easily use the method in this work to forecast the constraints using the momentum field measured from multiple surveys \citep{Park2000, Park2006, Okumura2014}.

We also use our formalism to investigate whether peculiar velocities can provide information on common extensions to the cosmological model and look at systematic effects in the modelling that have the potential to bias constraints from future peculiar velocity surveys. Over the course of this work, we will revisit the predictions from previous works and update them to more accurately reflect the current state of some of the future surveys. We aim to emphasise the cosmological benefit of peculiar velocity surveys, both separately and in combination, but highlight that with increased constraining power comes an increased need to consider potential modelling systematics.

The layout of the paper is as follows: In Sections~\ref{sec:power} and~\ref{sec:Fisher} we present our theoretical model for the Fisher matrix and power spectra measured from multiple surveys and tracers of the velocity and density fields. We also detail several extensions to our formalism to include and predict constraints on the effects of primordial non-Gaussianity, scale-dependent velocity and spatial bias, incorrect calibration of the zero-point and the $\gamma$ parameterisation for the linear growth rate. Sections~\ref{sec:data} and~\ref{sec:results} present the datasets considered in this work and the predicted constraints that can be recovered from them, both separately and in combination. In Section~\ref{sec:systematics} we look at the effects of the different systematics on constraints of the growth rate before concluding in Section~\ref{sec:conclusion}.

Throughout, unless otherwise stated, we use a flat, neutrinoless fiducial cosmology with cosmological constant, based on the results of \cite{Planck2015a}: $\Omega_{m}=0.3089$, $\Omega_{b}=0.0486$, $H_{0}=67.74$, $n_{s}=0.9667$ and $\sigma_{8}=0.8159$. With this cosmology our fiducial value for the growth rate is $f(z=0) = 0.524$.
 
\section{Model power Spectra} \label{sec:power}
The purpose of this work is to explore the cosmological information within the velocity and density fields as measured by (a combination of) peculiar velocity surveys. We will do this using a Fisher matrix method to look at how the information within the two-point correlations of these two fields. Within the Gaussian regime, where the Fisher matrix method is applicable, all the information in these fields is captured by just the two-point correlations. The first step then is to write down models for two-point correlations between measured velocity and density fields, in the form of the galaxy-galaxy, galaxy-velocity and velocity-velocity power spectra. These are not specific to the application presented in this work and could just as easily be used as models to fit against real data.

We adopt the redshift-space models of \cite{Koda2014}, briefly reiterating their derivation, before extending these to multiple tracers and including the effects of primordial non-Gaussianity, scale-dependent velocity and spatial bias, a zero-point calibration offset and looking at constraints on consistency tests of General Relativity using the well-known $\gamma$ parameterisation \citep{Linder2007}. The starting point is a model of the real-space dark matter density-density, density-velocity and velocity-velocity power spectra.

\subsection{Real-space matter power spectra} \label{sec:powerreal}
The power spectrum of some quantity can be written as the ensemble average over a volume $V$ of its products in Fourier space. In this work we are interested in two such quantities, the galaxy overdensity $\delta_{g}(\br)$, as inferred from the sky coordinates and redshifts of our galaxy sample, and the line-of-sight peculiar velocity $u(\br) = \bv(\br)\cdot\hat{\br}$. Using the redshifts to calculate the galaxy overdensity means that we need to deal with these quantities in redshift-space, including the effects of RSD on both the galaxy and velocity power spectra \citep{Seljak2011,Okumura2014}. However, the starting point for any redshift-space model is the clustering in real-space.

Other than the effects of redshift space distortions and galaxy bias, which will be added in the next section, the two-point clustering of the galaxy overdensity can be written in terms of the power spectrum of the matter overdensity $\delta_{m}(\bk)$. Using the definition of the power spectrum, this is $P_{mm}(\bk) = \langle \delta_{m}(\bk)\delta_{m}^{*}(\bk) \rangle / V$, which is the first of our real-space power spectra.

For the power spectra involving the line-of-sight velocity field, the exact choice of real-space power spectrum is largely one of convenience. In this work we choose to relate the Fourier transform of the line-of-sight peculiar velocity $u(\bk)$ to the velocity divergence via 
\begin{equation}
u(k,\mu) = -iaH(a)f\mu\theta(k)/k
\label{eq:lospv}
\end{equation}
and use the real-space velocity divergence power spectrum $P_{\theta\theta}(\bk) = \langle \theta(\bk)\theta^{*}(\bk) \rangle / V$. Here $\mu$ is the cosine of the angle between the observer's line-of-sight and the $k$-space vector. One could just as easily write a similar expression for the real-space line-of-sight velocity power spectrum, however using the real-space velocity divergence has two benefits. Firstly, storing and using the velocity divergence power spectrum allows us to change the value and form of the growth rate external of any code and without recomputing the real-space power spectra. Secondly, as can be seen from the linear continuity equation, the velocity divergence power spectrum should be equal to the matter power spectrum on linear scales (although on non-linear scales there will be some differences). This serves as a useful consistency check.

Finally, we can also define the cross-correlation between the matter overdensity field and the velocity divergence field in Fourier space, which gives the last of our real-space power spectra, $P_{m\theta}(\bk) = \langle \delta_{m}(\bk)\theta^{*}(\bk) \rangle / V$.

Real-space dark matter power spectra for the density field, velocity divergence field and their cross-correlation are obtained using the implementation of two-loop Renormalised Perturbation Theory (RPT; \citealt{Crocce2006a, Crocce2006b, Crocce2008}) found in the {\sc copter} numerical package \citep{Carlson2009}. This takes as input a normalised linear transfer function evaluated at redshift zero generated using {\sc camb} \citep{Lewis2000, Howlett2012}. 

By comparison to a suite of N-Body simulations, \cite{Carlson2009} found that two-loop RPT was able to recover the real-space density-density, density-velocity and velocity-velocity power spectra of a redshift $z=0$ $\Lambda$CDM universe to within $1\%$ up to $k=0.08\hompc$ and $\sim 8, 10$ and $15\%$ respectively up to $k=0.2\hompc$. They also found that this was typical when compared to a variety of different theoretical methods; most current models of the real-space power spectra become badly behaved on quasi-linear scales. For our forecasts the majority of the information on the linear growth factor comes from large scales, $k<0.1\hompc$, where RPT is reasonably reliable and as such we deem this approach sufficient for our needs.

\subsection{Redshift-space galaxy power spectra} \label{sec:powerred}
Under the assumption of a linear, stochastic galaxy bias and within the plane-parallel approximation, the redshift-space galaxy density field can be written in terms of the underlying matter density as
\begin{equation}
\delta_{g}(k,\mu) = D_{g}(k,\mu,\sigma_{g})[(b+f\mu^{2})\delta_{m}(k)]
\label{eq:delta}
\end{equation}
\citep {Dekel1999,Taylor2001,Burkey2004} where $b$ is the linear galaxy bias and the other parameters are as defined previously. $D_{g}$ parameterises the damping of the density field due to non-linear redshift space distortions. We adopt the Lorentzian damping model
\begin{equation}
D_{g}(k,\mu,\sigma_{g})  = \left[1+\frac{(k\mu\sigma_{g})^{2}}{2}\right]^{-1/2},
\end{equation}
where $\sigma_{g}$ is related to the pairwise velocity dispersion between pairs of galaxies. Although physically motivated $\sigma_{g}$ is generally a free parameter. It can be fixed by fitting the redshift-space power spectrum measured from simulations or marginalised over when fitting this model to data. In this work we treat it as a nuisance parameter and include the effect of marginalising over it on our constraints.

The redshift space version of Eq.~\ref{eq:lospv} can be written
\begin{equation}
u(k, \mu) = -iaH(a)f\mu\theta(k)D_{u}(k, \sigma_{u})/k
\label{eq:vel}
\end{equation} 
Redshift space distortions introduce a damping term $D_{u}$, which for this work is assumed to be a sinc function \citep{Koda2014}
\begin{equation}
D_{u}(k,\sigma_{u}) = \mathrm{sinc}(k\sigma_{u}).
\end{equation}
Again $\sigma_{u}$ is physically related to the non-linear motions of galaxies and the RSD they induce. We marginalise over it in all our Fisher matrix forecasts.  

The dark matter density and velocity divergence fields are intimately linked via the continuity equation, giving rise to the previously mentioned set of auto- and cross-power spectra $P_{mm}$, $P_{m\theta}$ and $P_{\theta \theta}$. Using Eqs.~\ref{eq:delta}, \ref{eq:vel}, and the cross-correlation coefficient, $r_{g}$, between the velocity and density fields as defined by \cite{Dekel1999}, these can be used to formulate expressions for the auto- and cross-power spectra measured from a set of galaxies in redshift-space,
\begin{align}
P^{AA}_{gg}(k,\mu) &= (\beta^{-2}_{A}+2r_{g,A}\beta^{-1}_{A}\mu^{2}+ \mu^{4})f^{2}D_{g,A}^{2}P_{mm}(k), \label{eq:pkbeg} \\
P^{AA}_{ug}(k,\mu) &= aH\mu k^{-1}(r_{g,A}\beta^{-1}_{A} + \mu^{2}) f^{2}D_{g,A}D_{u,A}P_{m\theta}(k), \\
P^{AA}_{uu}(k,\mu) &= (aH\mu)^{2}k^{-2} f^{2}D^{2}_{u,A}P_{\theta\theta}(k), \label{eq:pkend}
\end{align}
where $\beta_{A} = f/b_{A}$ and we have dropped the $k$ and $\mu$ dependence from $D_{g}(k,\mu)$ and $D_{u}(k,\mu)$ for brevity. The sub- and superscripts `A' and `AA' denote that these quantities are as measured from some galaxy sample/survey `A'. The necessity of including these to identify a particular sample will become apparent in Section~\ref{sec:powermulti} when we extend our formulae to look at multiple surveys. From the linear continuity equation, one might expect the velocity and density fields to be perfectly correlated, such that $r_{g}=1$ and it is unnecessary to include this parameter. However in the case of a stochastic and non-linear galaxy bias the correlation between the velocity and density fields may not be perfect. As such we include $r_{g}$ as a free parameter, with the fiducial value $r_{g}=1$, which is then marginalised over in our forecasts alongside $\sigma_{g}$ and $\sigma_{u}$. 

Eqs.~\ref{eq:pkbeg}-\ref{eq:pkend} relate the real-space power spectra from Section~\ref{sec:powerreal}, output by {\sc copter}, to those that would be measured from a peculiar velocity survey consisting of redshifts and line-of-sight peculiar velocity measurements for each galaxy. These expressions for the redshift-space galaxy-galaxy, galaxy-velocity and velocity-velocity power spectra, $P_{gg}(k,\mu)$, $P_{ug}(k,\mu)$ and  $P_{uu}(k,\mu)$, can be calculated by writing down the general formulae for the power spectra of $\delta_{g}(\bk)$ and $u(\bk)$ and using the equations in this and the previous section. Note that when doing so, it becomes apparent $P^{AA}_{ug}(k,\mu) = P^{AA}_{gu}(k,\mu)$.

This trio of models was adopted and tested by \cite{Koda2014} using the GiggleZ simulation \citep{Poole2015}. They find these models to be a good fit to the redshift space power spectra measured from the simulation for $k<0.2\hompc$, which is sufficient for the forecasts within this study.

\subsubsection{Parameter values} \label{sec:params}
Overall, Eqs.~\ref{eq:pkbeg}-\ref{eq:pkend} depend on the growth rate, $f$, the survey specific value of the galaxy bias, $b_{A}$, and the three nuisance parameters $r_{g,A}$, $\sigma_{g,A}$ and $\sigma_{u,A}$. For our forecasts, on top of our fiducial cosmology, we assume a value of $f=\Omega_{m}^{0.55}=0.524$, which matches the prediction from General Relativity \citep{Linder2007}. Based on the best-fitting values of the RSD damping parameters and cross-correlation coefficient found by \cite{Koda2014}, we adopt values of $r_{g}=1.0$, $\sigma_{g}=4.24\mpcoh$ and $\sigma_{u}=13.0\mpcoh$. The exact values used for these nuisance parameters are expected to have little impact on the predicted growth rate constraints. For multiple tracers, as described in the next section, we marginalise over multiple sets of these nuisance parameters, however these are given the same fiducial values. As found by \cite{Koda2014}, these values have a slight dependence on halo mass, reflecting how different halo masses and galaxy populations undergo non-linear RSD in different ways. Different surveys, containing different samples of galaxies, could be expected to have different best-fitting values for our non-linear RSD parameterisation but as the forecasts are expected to be robust to the exact values used, we use the same fiducial value for each survey.

\subsection{Power spectra for multiple tracers} \label{sec:powermulti}
Now we turn to the case of two surveys, A and B, each with their own set of redshifts and peculiar velocities. In addition to the three power spectra in Eq.~\ref{eq:pkbeg}-\ref{eq:pkend}, there are three unique power spectra associated with the correlation between two galaxies from survey B ($P^{BB}_{gg}$, $P^{BB}_{ug}$ and $P^{BB}_{uu}$) and 4 from cross-correlations of the velocities and densities measured from the two samples ($P^{AB}_{gg}$, $P^{AB}_{ug}$, $P^{BA}_{ug}$ and $P^{AB}_{uu}$). In total there are 10 distinct power spectra that can be formulated. However, in our fiducial model, we assume that both sets of galaxies will trace the underlying velocity field in the same way, i.e., there is no scale-dependent velocity bias, or rather the velocity bias is unity on all scales. In this case $P^{AB}_{ug} = P^{BB}_{ug}$, $P^{BA}_{ug} = P^{AA}_{ug}$, and $P^{AA}_{uu} = P^{AB}_{uu} = P^{BB}_{uu}$. Hence the only cross-spectra of interest between the two tracers is
\begin{align}
P^{AB}_{gg}(k,\mu) &= (\beta^{-1}_{A}\beta^{-1}_{B}+(\beta^{-1}_{A}r_{g,A} + \notag \\
& \qquad \beta^{-1}_{B}r_{g,B})\mu^{2}+ \mu^{4})f^{2}D_{g,A}D_{g,B}P_{mm}(k).
\label{eq:pkmult}
\end{align}
This equation, combined with Eqs.~\ref{eq:pkbeg}-\ref{eq:pkend} for surveys A and B and the set of equalities given above covers all possible two point correlations between the density and velocity fields measured from two surveys. Overall, our fiducial model has 6 unique spectra. The assumption that both surveys trace the velocity field in the same way will be relaxed in the following section. $\beta_{B}=f/b_{B}$ where $b_{B}$ is the bias for survey `B'. This does not have to be equal to $b_{A}$. Indeed, as pointed out by \cite{Mcdonald2009b}, because different surveys trace the same value of $f$ but have potentially different values for the galaxy bias, the use of multiple samples with different biases can be very effective at breaking the degeneracy between the growth rate and bias on linear scales. Note that we have used different subscripts for the cross-correlation coefficients, $r_{g}$, and non-linear RSD damping terms $D_{g}$ and $D_{u}$, reflecting the fact that even though we assume the same fiducial values for all surveys, these are actually distinct for each survey and are marginalised over separately in the subsequent parts of this work.

\begin{figure*}
\centering
\subfloat{\includegraphics[width=0.33\textwidth]{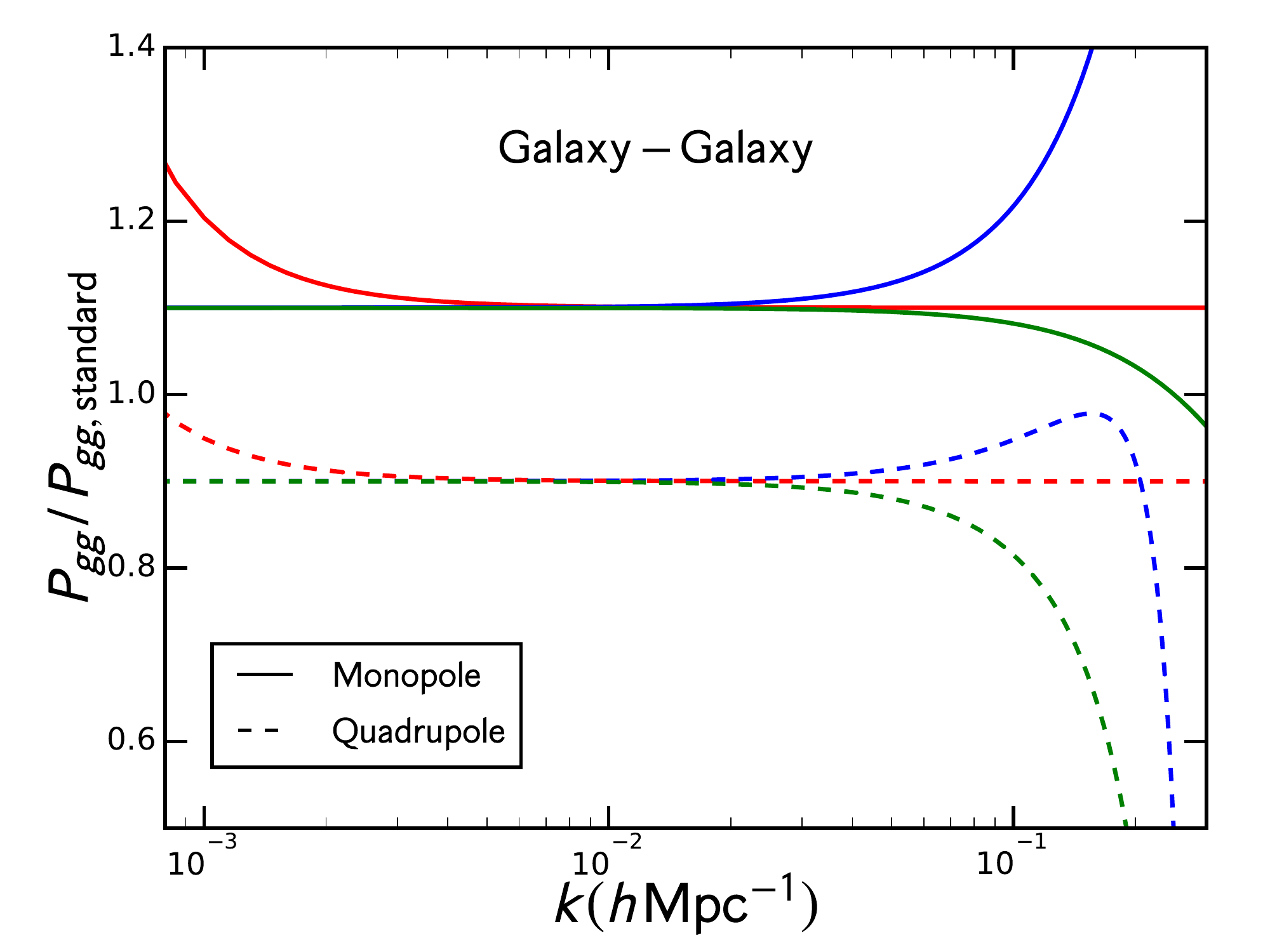}}
\subfloat{\includegraphics[width=0.33\textwidth]{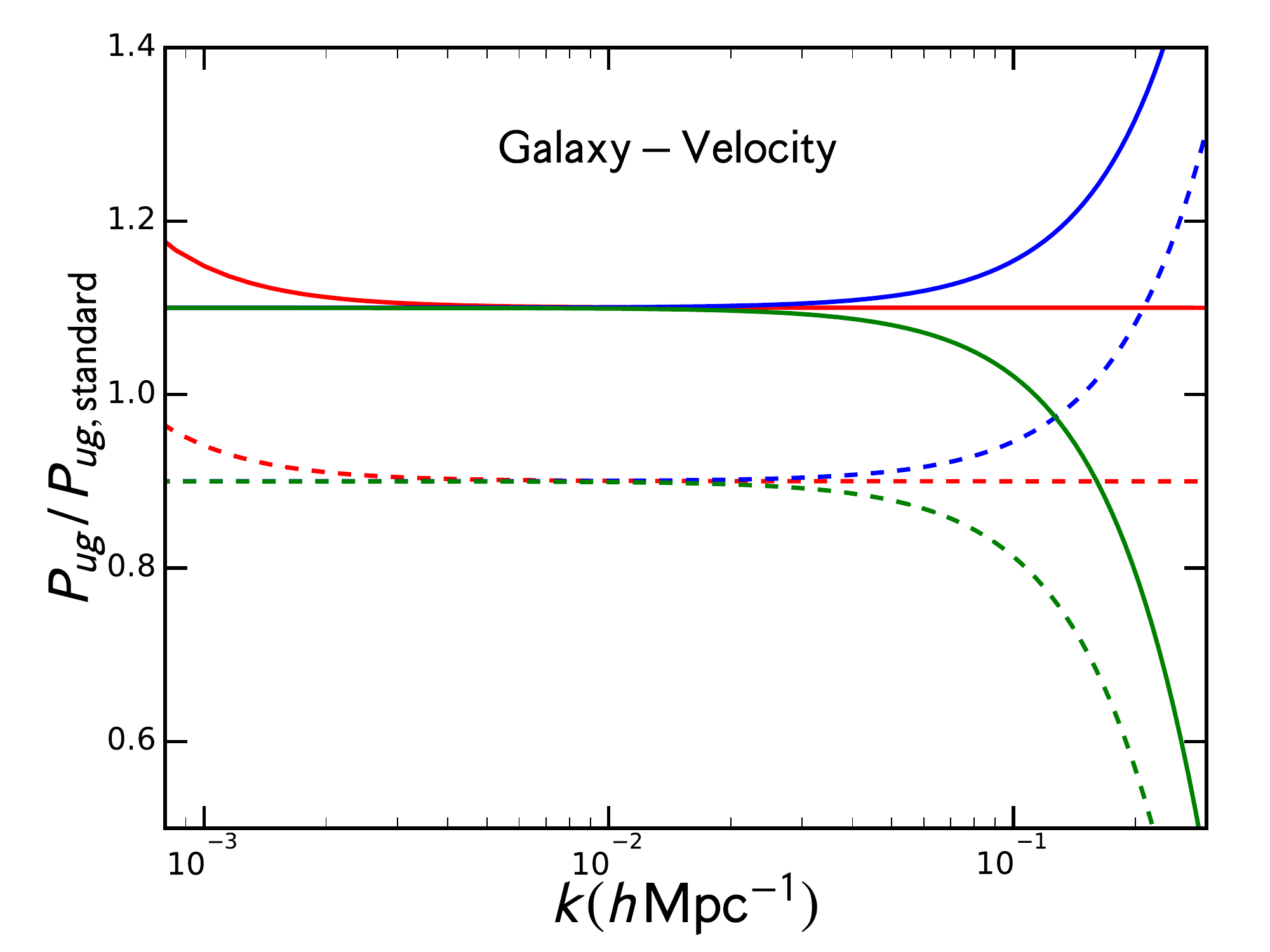}}
\subfloat{\includegraphics[width=0.33\textwidth]{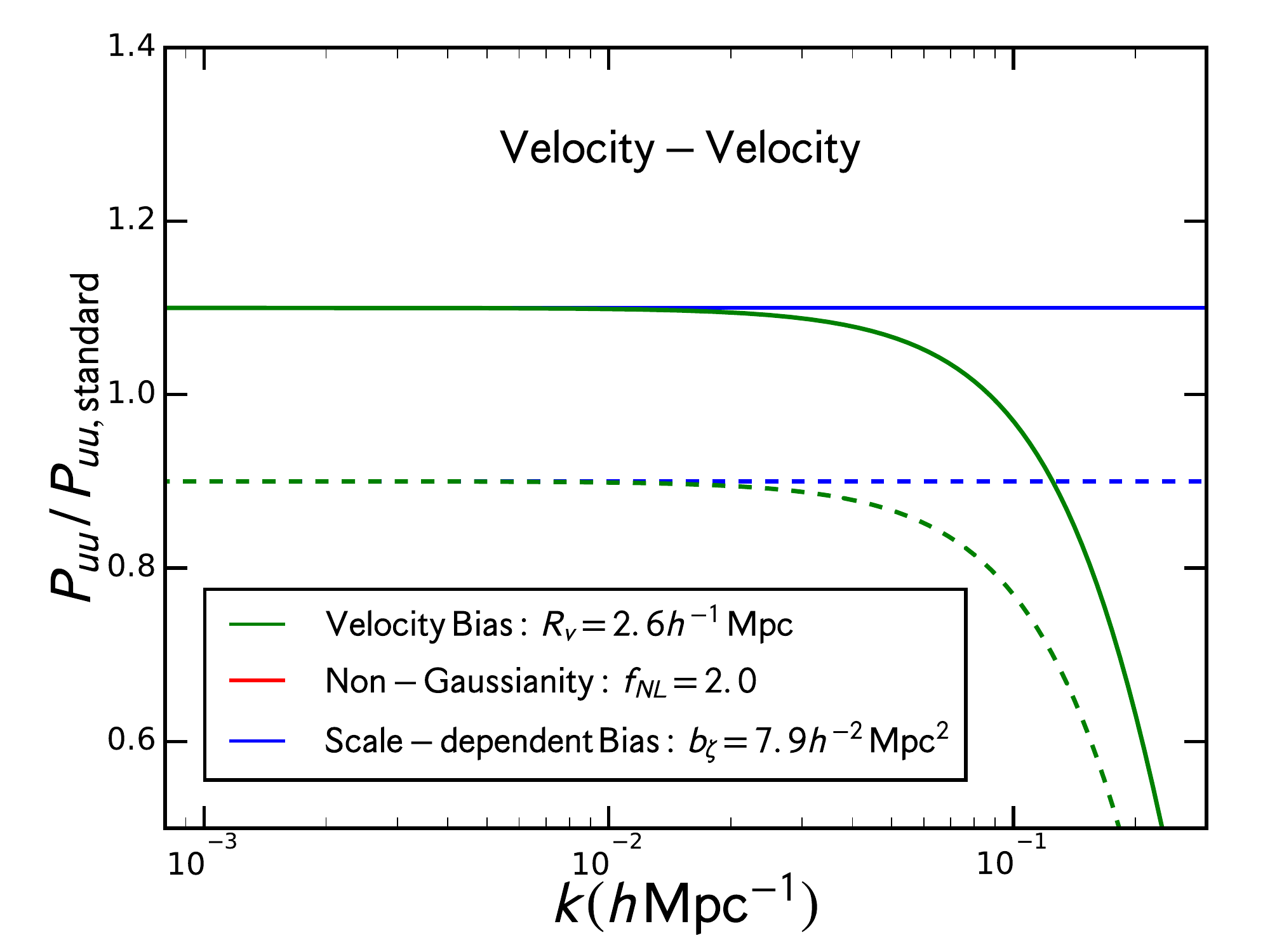}}
  \caption{An illustrative example of the effect of scale-dependent spatial bias (blue), velocity bias (green) and primordial non-Gaussianity (red) on the monopole and quadrupole of the galaxy-galaxy, galaxy-velocity and velocity-velocity power spectra. The moments of the three spectra are defined in Eq.~\ref{eq:pkmoments} and are used to remove the $\mu$ dependence from the models. In each case we plot the ratio of the power spectra with the model extension over the standard model, Eqs.~\ref{eq:pkbeg}-\ref{eq:pkend} (with the monopole/quadrupole shown as solid/dashed lines and offset by $\pm 0.1$ for clarity). We use the fiducial parameter values given in Section~\ref{sec:powerred} with galaxy bias, primordial non-Gausianity, scale-dependent spatial and velocity bias characterised by $\beta=0.435$, $f_{NL}=2.0$, $b_{\zeta}=7.9h^{-2}\,{\rm Mpc}^{2}$ and $R_{v}=2.6\mpcoh$ respectively. These are the same as those adopted for our systematic tests on the TAIPAN survey (c.f. Section~\ref{sec:datataipan}). Primordial non-Gaussianity modifies the galaxy bias on large scales (small $k$) and adds power for our choice of parameters, as shown by the upturn in the red lines at small $k$. The scale-dependent spatial bias generally increases the power on small scales, shown as an upturn in the blue line at high $k$, except for the quadrupole of the galaxy-galaxy power spectrum. This results from the particular combination of $\beta$ and $\mu$ terms in Eq.~\ref{eq:pkbeg}. The velocity bias significantly reduces the small scale power for all three spectra. For the velocity-velocity power spectrum the red and blue lines are constantly equal to the fiducial model as our model velocity power spectrum is independent of galaxy bias (and hence primordial non-Gaussianity and scale-dependent spatial bias).}
  \label{fig:pk_example}
\end{figure*}

\subsection{Extensions to the standard formalism}
In this study we wish to look at the ability of peculiar velocity surveys to provide constraints on parameters beyond our fiducial model that may produce a measurable signal or quantifiable systematic effect on the scales probed by the velocity field. How these are incorporated into our models is given in this section.

An illustrative example of the effects of some of the extensions to our models, for reasonable parameter choices (we use the same values as for the TAIPAN dataset detailed in Section~\ref{sec:datataipan}), is given in Fig.~\ref{fig:pk_example}. In this Figure we plot the ratio of the monopolar and quadrupolar moments of the galaxy-galaxy, galaxy-velocity and velocity-velocity power spectra with and without the effects of primordial non-Gaussianity, scale-dependent spatial galaxy bias and velocity bias as separate lines. The moments of the galaxy-galaxy power spectra are defined as
\begin{equation}
P^{\ell}_{gg}(k) = \frac{2}{2\ell+1}\int_{0}^{1} d\mu P_{gg}(k,\mu) \mathcal{P}^{\ell}(\mu)
\label{eq:pkmoments}
\end{equation}
where $\mathcal{P}^{\ell}(\mu)$ are the Legendre polynomials and similar expressions are used for the galaxy-velocity and velocity-velocity power spectrum moments. These moments are used to remove the angular $\mu$ dependence of the power spectrum and allow the effects of the different model extensions to be shown as only a function of scale. In all cases, the regimes where the model extension has no effect should show as a constant value of $1.0$ (although we have offset the monopole and quadrupole by $\pm0.1$ for clarity.)

\subsubsection{Primordial non-Gaussianity} \label{sec:fnlmodel}
Under the local ansatz, primordial non-Gaussian perturbations arising from certain inflationary models have a Bardeen potential quadratic about the Gaussian field $\phi$, i.e., $\Phi=\phi+f_{NL}(\phi^{2}-\langle\phi^{2}\rangle)$. The parameter $f_{NL}$ quantifies the strength of the deviation from Gaussianity. These same non-Gaussian perturbations introduce a scale dependent addition to the galaxy bias \citep{Dalal2008,Matarrese2008}, which can in turn be used to constrain $f_{NL}$ and hence place limits on the type of inflation that took place in the early universe. The change in the galaxy bias takes the form
\begin{equation}
b_{tot} = b+\Delta b
\end{equation}
where
\begin{equation}
\Delta b = 3f_{NL}(b-1)\frac{\delta_{c}\Omega_{m,0}H^{2}_{0}}{k^{2}T(k)\bar{D}(z)c^{2}},
\label{eq:biasfnl}
\end{equation}
$\delta_{c}=1.686$ is the critical density threshold for spherical collapse, $T(k)$ is the matter transfer function normalised to one as $k\rightarrow 0$ and $\bar{D}(z)$ is the linear growth factor normalised to unity at the present day.
The effect of primordial non-Gaussianity for $f_{NL}=2.0$, which is within the current constraints from \cite{Planck2015b}, is shown in Fig.~\ref{fig:pk_example}. It is an increase in the galaxy-galaxy and galaxy-velocity power spectra at low-$k$ that increases as one goes to larger scales. It is most apparent for the monopole of the galaxy-galaxy power spectrum, which has the strongest dependence on the galaxy bias. Our model velocity power spectrum is unaffected by local primordial non-Gaussianity of the form used here as it has no dependence on galaxy bias.

Being partially degenerate with the galaxy bias, and only really having an effect on the largest scales, means that constraining $f_{NL}$ using a single galaxy survey can be difficult, though this has been done before \citep{Ross2013}. However because $f_{NL}$ is very sensitive to the bias the use of multiple-tracers can have a large impact on constraints \citep{Seljak2009,Mcdonald2009b}. Peculiar velocities can be also used to place constraints on $f_{NL}$, both via a comparison of the measured and reconstructed velocity fields under the assumption of some bias model \citep{Ma2013} and in combination with density field measurements by alleviating some of the degeneracy between $f_{NL}$ and other parameters.

It is interesting to see whether current or upcoming peculiar velocity surveys may be able to place constraints on $f_{NL}$. In this study this is done by modifying the galaxy bias, and hence the value of $\beta$, of each galaxy sample according to Eq.~\ref{eq:biasfnl}. We use the same linear matter transfer function, output from {\sc camb}, as was used to generate the real-space power spectra in Section~\ref{sec:powerreal}. Different surveys will experience the effects of a fixed value of $f_{NL}$ differently due to their different galaxy biases. For all our forecasts we adopt a value $f_{NL}=2.0$ motivated by the results of \cite{Planck2015b}.

\subsubsection{$\gamma$ parameterisation} \label{sec:gamma}
The $\gamma$ parameterisation of \cite{Linder2007},
\begin{equation}
f(z) = \Omega_{m}(z)^{\gamma}
\end{equation}
is commonly used to test the consistency of measurements of the growth rate with General Relativity, where $\gamma=0.55$. Because of its simplicity this parameterisation has seen wide usage, however it is difficult to relate a particular measured value of $\gamma$ to some modified gravity model, in particular because of its inability to model any scale dependence in the growth rate. For this reason, in this study we do not adopt this parameter into our standard model. However, we do include it as an extension as it is of interest to see what constraints can be placed on $\gamma$ by both current and future peculiar velocity surveys when scale independence is assumed.

A fully self-consistent treatment of this parameter would include the fact that different values of $\gamma$, or rather different modified gravity theories, will change the shape of the matter power spectrum, its present day amplitude (usually parameterised by $\sigma_{8}$, the variance of the linear matter field in spheres of radius $8\mpcoh$) and the growth rate of structure. However, the shape of the power spectrum is very strongly constrained by the CMB \citep{Planck2015a}, and hence, in practice, most measurements of the growth rate simply constrain the parameter combination $f\sigma_{8}$ (see \citealt{Song2009} for an examination of why this combination in particular is usually measured). So whilst changes in the shape of the power spectrum can be neglected, it is still important to also include the fact that the value of $\sigma_{8}$ will depend on the value of $\gamma$.  

We follow the method of \cite{Howlett2015} and account for this effect by scaling the value of $\sigma_{8}$ measured under the assumption of GR back to some suitably high redshift (for example the redshift of recombination $z*$) using the linear growth factor, then scaling it forward under the new cosmology, i.e., at scale factor $a=1$
\begin{equation}
f\sigma_{8} = \Omega_{m,0}^{\gamma}\sigma_{8}\frac{D_{gr}(a*)}{D_{gr}(1)}\frac{D_{\gamma}(1)}{D_{\gamma}(a*)}
\end{equation}
where
\begin{equation}
D_{gr}(a) = \frac{H(a)}{H_{0}}\int_{0}^{a}\frac{da'}{a'^{3}H(a')^{3}},
\end{equation}
\begin{equation}
\frac{D_{\gamma}(1)}{D_{\gamma}(a*)} = \text{exp}\left[ \int_{a*}^{1} \Omega_{m}(a')^{\gamma} d\text{ln}a' \right]
\end{equation}
and
\begin{equation}
H(a) = H_{0}E(a) = H_{0}\sqrt{\frac{\Omega_{m,0}}{a^{3}}+\frac{(1-\Omega_{m,0}-\Omega_{\Lambda,0})}{a^{2}}+\Omega_{\Lambda,0}}
\end{equation}

\subsubsection{Scale-dependent spatial and velocity bias} \label{sec:velbias}
On large scales galaxies are assumed to be linearly biased with respect to the dark matter. However, it has long been established that scale-dependent galaxy bias exists on smaller scales (for recent studies see \citealt{Scoccimarro2004,Mcdonald2009a,Seljak2011,Chan2012,Baldauf2013,Saito2014} and references therein). Several studies have looked at the impact of this on measurements of the growth rate \citep{Smith2007,Poole2015,Amendola2015}. To overcome this, measurements of the growth rate typically include higher order bias terms in the models \citep{Beutler2014,Gil-Marin2015,Howlett2015} or truncate their fits at scales where the bias is expected to remain linear \citep{Beutler2012, Samushia2014}. In addition to being a potential source of systematic error, measuring the scale-dependence of the galaxy bias can give interesting insight into the relationship between galaxies and their host dark matter halos.

In the context of cosmological measurements it is usually assumed that the velocity divergence measured from a set galaxies exactly follows the underlying velocity divergence field. That is, $\theta_{g}=b_{v}\theta_{m}$ where $b_{v}=1$. However, several studies \citep{Desjacques2008, Desjacques2010a, Biagetti2014, Baldauf2015} have presented arguments of how the velocities of peaks in the density field may be statistically biased with respect to the underlying velocity field in a scale dependent way. 

In recent years many studies have attempted to measure this velocity bias and its effect by comparing the velocity divergence power spectrum measured from simulated halos and the corresponding dark matter field \citep{delaTorre2012,Elia2012,Jennings2015,Zheng2015}, however the magnitude of this effect remains largely uncertain due to numerical resolution issues and difficulties in measuring the velocity divergence power spectrum. Regardless, most studies agree that the effect of the velocity bias is small for $k<0.1\hompc$, and hence one could argue that it remains unimportant for current measurements of the growth rate of structure from peculiar velocities, which relies primarily on the information from linear scales. Nonetheless, as the constraining power of peculiar velocity surveys increases it is of interest to investigate the effect this unknown parameter could have of constraints of the growth rate. This is especially true for scale-dependent growth rate measurements, where the scale dependence of the velocity bias can be misconstrued as a signature of modified dark energy or gravity models.

To investigate the possible effects of scale-dependent spatial and velocity bias, we adopt the following model. Using the peaks approach, \cite{Desjacques2010a} show that the spatial bias of peaks in the density field follows
\begin{equation}
b_{sd} = b + b_{\zeta}k^{2},
\end{equation}
where $b$ is the standard linear bias and $b_{sd}$ is the new scale-dependent bias. Similarly, the velocity bias has a $k^{2}$ dependence of the form
\begin{equation}
b_{v} = 1-R^{2}_{v}k^{2}.
\label{eq:velbias}
\end{equation}
$b_{\zeta}$ and $R^{2}_{v}$ are the normalisation of the scale dependence, both of which depend on the characteristic scale and mass of the tracers in question. Different values for these parameters for different datasets will be used in this study. A reasonable choice of values can be theoretically motivated for a given mass $M$ using the spectral moments of the matter power spectrum, $\sigma_{n}(R)$, smoothed within a Gaussian filter of some radius, $R$. From \cite{Bardeen1986},
\begin{equation}
\sigma_{n}^{2}(R) = \frac{1}{2\pi^2} \int_{0}^{\infty} dk k^{2(n+1)} P(k)W^{2}(k,R),
\label{eq:specmo}
\end{equation}
where
\begin{equation}
W^{2}(k,R) = e^{-k^{2}R^{2}},\,M=(2\pi)^{3/2}\Omega_{m,0}\rho_{c}R^{3}
\end{equation}
and $\rho_{c}$ is the critical density. This is very similar to the expression used to define the well-known $\sigma_{8}$ parameter, with $n=0$ and a spherical tophat window of radius $8\mathrm{Mpc}$, rather than a Gaussian window of radius $R$. From the spectral moments we can also define $\psi = \sigma^{2}_{1}/(\sigma_{0}\sigma_{2})$. Then we can calculate the scale-dependent normalisations as
\begin{align}
b_{\zeta} &= \frac{1}{\sigma_{2}}\frac{\bar{u} - \psi\nu}{1 - \psi^{2}}, \\
R_{v} &= \frac{\sigma_{0}}{\sigma_{1}}. \label{eq:Rv}
\end{align}
In addition to the spectral moments, $b_{\zeta}$ also depends on the peak height $\nu = \delta_{c}/\sigma_{0}$ and mean curvature $\bar{u}$. $\delta_{c}=1.686$ is the critical density threshold for spherical collapse. The mean curvature is also a function of the peak height and $\psi$, albeit a rather complex one, requiring numerical integration to solve fully. As the expressions are lengthy, we do not reproduce them here, but they can be readily found within the Appendices of \cite{Desjacques2010b} (Eqs. A59-A60 and the accompanying text). Approximate fitting functions for the required integrals for peaks with height $v >> 1$ can be found in \cite{Bardeen1986} (Eqs. 4.4 and 4.5).

\begin{figure}
\centering
\includegraphics[width=0.5\textwidth]{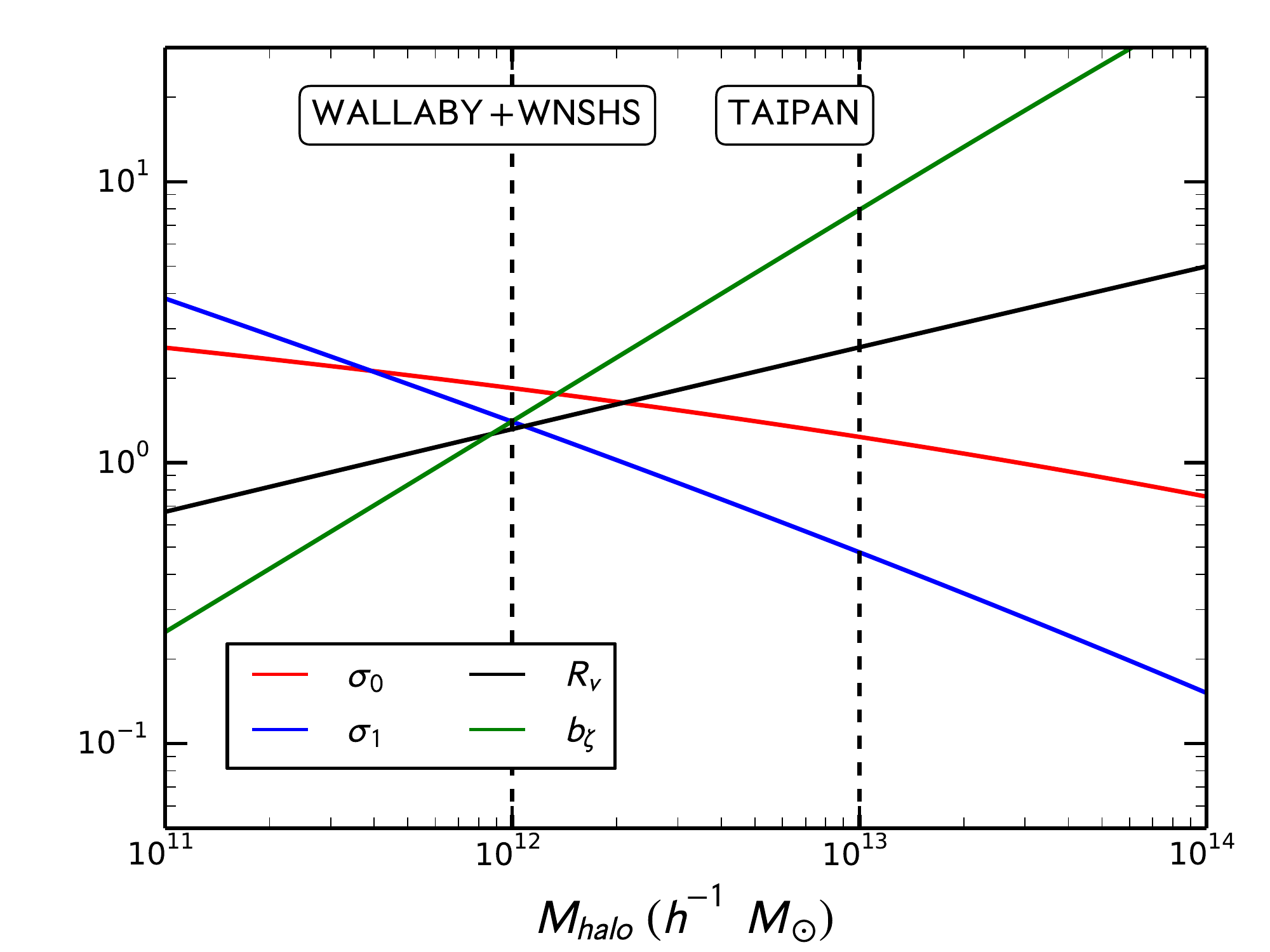}
  \caption{The spectral moments, $\sigma_{0}$ and $\sigma_{2}$, and resulting normalisation of the scale-dependent spatial and velocity biases, $b_{\zeta}$ and $R_{v}$, for a range of halo masses at $z=0$ for our fiducial cosmology. Also identified is the typical halo mass and normalisations adopted for the TAIPAN and WALLABY+WNSHS datasets.}
  \label{fig:rvel}
\end{figure}

The spectral moments, $b_{\zeta}$ and $R_{v}$ are plotted in Fig.~\ref{fig:rvel} for a range of halo masses. These are calculated by solving our Eqs.~\ref{eq:specmo}-\ref{eq:Rv} and Eqs. A59-A60 from \cite{Desjacques2010b}. We also highlight the typical halo masses we assume for the TAIPAN and WALLABY+WNSHS datasets. The corresponding values of $b_{\zeta}$ and $R_{v}$ are given in Section~\ref{sec:data}. \cite{Elia2012} found that the form of the scale-dependent spatial and velocity bias was well matched to that measured from simulations.

In the presence of velocity bias the auto- and cross-power spectra for a given sample are modified from those given in Eq.~\ref{eq:pkbeg}-\ref{eq:pkend} to:
\begin{align}
P^{AA}_{gg}(k,\mu) &= (\beta^{-2}_{A}+2b_{v,A}r_{g,A}\beta^{-1}_{A}\mu^{2}+b^{2}_{v,A}\mu^{4}) \notag \label{eq:pkbvbeg}\\ 
& \qquad \qquad \qquad \qquad \qquad \qquad f^{2}D_{g,A}^{2}P_{mm}(k), \\ 
P^{AA}_{ug}(k,\mu) &= aH\mu k^{-1}(r_{g,A}\beta^{-1}_{A} + b_{v,A}\mu^{2})b_{v,A} \notag \\
& \qquad \qquad \qquad \qquad \qquad f^{2}D_{g,A}D_{u,A}P_{m\theta}(k), \\
P^{AA}_{uu}(k,\mu) &= (aH\mu)^{2}k^{-2} b^{2}_{v,A}f^{2}D^{2}_{u,A}P_{\theta\theta}(k).
\end{align}
The velocity bias has an effect on all three power spectra. These can be calculated in the same way as Eq.~\ref{eq:pkbeg}-\ref{eq:pkend}, but replacing $u_{g}(k,\mu) = b_{v}u_{m}(k,\mu)$, i.e., the line-of-sight peculiar velocity measured from a galaxy field is no longer exactly equal to that of the underlying dark matter. This is equivalent to adding $b_{v}$ to Eq.~\ref{eq:vel}. Similar expressions are obtained for a second sample, $B$. The scale-dependent spatial bias is absorbed into the $\beta$ parameter as $\beta\rightarrow (1/\beta + b_{\zeta}k^{2}/f)^{-1}$. Also note that in the presence of velocity bias several of the cross-spectra between the two fields can no longer be written in terms of the correlation between two points on the same field as each survey may trace the velocity field differently, i.e., $P^{AB}_{ug} \neq P^{BB}_{ug}$. In particular, the cross-spectra between the two fields now take the form
\begin{align}
P^{AB}_{gg}(k,\mu) &= (\beta^{-1}_{A}\beta^{-1}_{B}+(\beta^{-1}_{A}r_{g,A}b_{v,A}+\beta^{-1}_{B}r_{g,B}b_{v,B})\mu^{2} \notag \\
& \qquad + b_{v,A}b_{v,B}\mu^{4})f^{2}D_{g,A}D_{g,B}P_{mm}(k), \\
P^{AB}_{ug}(k,\mu) &= aH\mu k^{-1}(r_{g,B}\beta^{-1}_{B} + b_{v,B}\mu^{2})b_{v,A} \notag \\
& \qquad \qquad \qquad \qquad \qquad f^{2}D_{g,B}D_{u,A}P_{m\theta}(k), \\
P^{BA}_{ug}(k,\mu) &= aH\mu k^{-1}(r_{g,A}\beta^{-1}_{A} + b_{v,A}\mu^{2})b_{v,B} \notag \\
& \qquad \qquad \qquad \qquad \qquad f^{2}D_{g,A}D_{u,B}P_{m\theta}(k), \\
P^{AB}_{uu}(k,\mu) &= (aH\mu)^{2}k^{-2}b_{v,A}b_{v,B}f^{2}D_{u,A}D_{u,B}P_{\theta\theta}(k). \label{eq:pkbvend}
\end{align}
Although these look very similar to the previous expressions and to each other, there are subtle differences in the exact combination of the two velocity biases when correlating fields from different surveys. The effect of scale-dependent spatial and velocity bias on the power spectra for $b_{\zeta}=7.6$ and $R_{v}=2.6\mpcoh$ is shown in Fig.~\ref{fig:pk_example}, and is a change in the galaxy-galaxy, galaxy-velocity and velocity-velocity power spectra on small scales.

\subsubsection{Zero-point offsets} \label{sec:zp}
Any determination of the peculiar velocities of a sample of galaxies requires calibration of the zero-point of the astrophysical relation used to infer the galaxy's true distance. In the case of the Tully-Fisher relation or another relation relating the source magnitude to some intrinsic property, this is the reference magnitude at which the peculiar velocity is known to be zero. For the Fundamental Plane relation it is some reference size. Without calibration of the zero-point only relative velocities between objects in the sample can be inferred. 

The zero-point is typically calibrated during fitting of the astrophysical relationship, however such a calibration can carry considerable uncertainty. For hemispherical surveys, the zero-point can be calibrated to reasonable accuracy such as to rule out biases in the zero-point due to large dipolar motions in the velocity field, but large offsets between the true and measured zero-points are still possible in the presence of a monopole. This monopole can arise due to sample variance and is equivalent to a change in the measured expansion rate due to local inhomogeneities. Whilst full-sky surveys are to be less impacted by the presence of a dipole (although due to the reality of uneven sky coverage in any survey, the advantage of being full-sky does not preclude this), the effects of sample variance are only expected to diminish as the volume covered by a survey becomes sufficiently large. \cite{Johnson2014} identify the potential for biases from zero-point offsets and demonstrate a neat way of marginalising over this in their measurements of the growth rate from 6dFGSv data.

A zero-point offset manifests as some net peculiar velocity, $u^{ZP} \neq 0$. This in turn gives an additive term to the measured velocity power spectrum
\begin{equation}
P^{AA}_{uu} \rightarrow P^{AA}_{uu} + \frac{(\sigma^{A}_{ZP})^{2}}{\bar{n}^{A}_{u}},
\end{equation}
i.e., the zero-point offset acts as a shot-noise contribution to the power spectrum. $\sigma_{ZP}$ is the error in the zero-point calibration whilst $\bar{n}_{u}$ is the number density of peculiar velocity tracers. We will revisit this latter term in Section~\ref{sec:Fisher}. This shot-noise term can be easily seen if the power in the measured line-of-sight peculiar velocity, $u^{M}$, is written in terms of the true velocity, $u^{T}$, plus some velocity introduced by the zero-point offset, $u^{M} = u^{T}+u^{ZP}$, and one re-performs the derivation of a power spectrum from some observable as in Section~\ref{sec:powerreal}.

For simplicity one could adopt a constant value of $\sigma_{ZP}$ for each survey and investigate the resultant effects on the growth rate constraints. However, in this study we use a more realistic approach. Errors in the zero-point calibration are typically given as a constant error in the reference magnitude, $\sigma_{m}$. A constant error in the reference magnitude corresponds to a redshift dependent error in the peculiar velocity, such that the additive term to the velocity power spectrum is also redshift dependent. Conveniently, \cite{Hui2006}, \cite{Davis2011} and \cite{Johnson2014} provide expressions for converting between apparent magnitudes and peculiar velocities
\begin{equation}
\sigma_{ZP} = \frac{c\,\mathrm{ln} 10}{5}\left(1 - \frac{c\,(1+z)}{H(z)r(z)}\right)^{-1}\sigma_{m}
\label{eq:zperr}
\end{equation}
where $r(z)$ is the comoving distance and $c$ is the speed of light. When looking at the systematic effects of the zero-point offset, we use this expression to convert our constant zero-point offset into a redshift dependent addition to the velocity power spectrum for each dataset. When calculating the Fisher matrix, this means it can simply be treated in a similar fashion to the redshift dependent shot-noise terms shown in Section~\ref{sec:fishernoise}.

In our study we choose a suitable value for the zero-point offset as follows. By fitting the 6dFGSv data within a ``great circle" between $-20^{\circ} \leq \delta \leq 0^{\circ}$, \cite{Springob2014} find a statistical error on the zero-point offset of $\sigma_{m}=0.015\,\mathrm{dex}$. This encompasses the effect of a dipole in the velocity field. In the presence of a monopole though, we could expect a larger offset in the zero-point. This is investigated by \cite{Scrimgeour2016} using a suite of mock simulations that mimic the selection function of the 6dFGSv sample. The use of multiple realisations allows for characterisation of the cosmic variance. \cite{Scrimgeour2016} find that the rms variance in the zero-point due to cosmic variance is $\sim0.02\,\mathrm{dex}$. Overall then, in order to be conservative and predict the largest systematic bias we could reasonably expect on the growth rate constraints from a zero-point offset, we take as our value $\sigma_{m}=0.05\,\mathrm{dex}$. This would correspond to a zero-point offset $\sim 2\sigma$ times greater than that found combining both the statistical and systematic uncertainties from the 6dFGSv sample. For the next generation surveys we consider we would expect the zero-point offset arising from a monopole to be no greater than that for 6dFGSv, as the cosmological volume covered will be equal or larger.

\subsection{Summary}

This section has presented an in-depth overview of the models we use for the power spectra between galaxy overdensities and peculiar velocities in this work. The galaxy-galaxy, galaxy-velocity and velocity-velocity power spectra should be readily measurable from current and future peculiar velocity surveys. In order to predict how such measurements correspond to constraints on the growth rate and other cosmological parameters, the models from this section will be used as input into the Fisher matrix calculation presented in the next section. We re-emphasise the point that these models are general however, and do not have to be used solely for Fisher matrix forecasts. As shown in \cite{Koda2014} they also correspond well to measurements from simulations and so could be used to estimate cosmological parameters from actual data.

To summarise, here are the main points of this section:
\begin{itemize}
\item{The starting point of the galaxy-galaxy, galaxy-velocity and velocity-velocity power spectra are a set of real-space power spectra generated using the {\sc copter} numerical code. This code generates non-linear versions of the real-space matter power spectrum, $P_{mm}$, velocity divergence power spectrum $P_{\theta\theta}$ and the cross spectrum $P_{m\theta}$.}
\item{These real-space spectra can then be turned into a set of redshift-space spectra as would be observed by some survey using Eqs.~\ref{eq:pkbeg}-\ref{eq:pkend}. For this we need an estimate of the properties of the galaxy sample within that survey, in particular the bias, $b$, the cross-correlation coefficient $r_{g}$, and non-linear RSD parameters $\sigma_{g}$ and $\sigma_{u}$. For this work we adopt suitable fiducial values for these parameters (c.f. Sections~\ref{sec:params} and~\ref{sec:data}), but ultimately treat them as unknowns and marginalise over them for our predictions.}
\item{Our models are extended to multiple surveys by adding a second set of Eqs.~\ref{eq:pkbeg}-\ref{eq:pkend}, \textit{with different survey parameters}. We then have to consider the cross-correlation between the two surveys as per Eq.~\ref{eq:pkmult} and the accompanying text.}
\item{There are also several model extensions beyond our fiducial model that we are interested in, namely primordial non-Gaussianity, scale-dependent spatial and velocity bias, and an offset in the zero-point. Sections~\ref{sec:fnlmodel}-\ref{sec:zp} describe these effects and how we model them by using alternative versions of Eqs.~\ref{eq:pkbeg}-\ref{eq:pkend}, such as Eqs.~\ref{eq:pkbvbeg}-\ref{eq:pkbvend}. In all cases we can recover our standard models by choosing suitable values for the additional parameters we introduce.}
\end{itemize}

\section{Fisher matrix} \label{sec:Fisher}

The goal of this work is to look at the cosmological information available within (multiple) current and upcoming peculiar velocity surveys. We do this using the well-known statistical Fisher information matrix method.
The Fisher information matrix gives the information content of an observable with respect to some underlying parameters $\boldsymbol{\lambda}$. In our case we are interested in the information on the growth rate and other cosmological parameters carried by the observables $\delta_{g}(\br)$ and $u(\br)$. In fact in this study we use their Fourier space equivalents, which by definition should carry the same amount of information.

The Fisher information with respect to some parameter is given by
\begin{equation}
\mathcal{F}(\lambda) = -\left\langle \frac{\partial^{2}\mathcal{L}}{\partial\lambda^{2}} \right\rangle,
\end{equation}
where we take the expectation of the second derivative of the likelihood, $\mathcal{L}$, with respect to that particular combination of parameters. The likelihood is generally calculated based on some model or measurements, which are dependent on the parameters we are interested in.

The Fisher information for a single parameter can be extended to multiple covariant parameters into the Fisher information matrix
\begin{equation}
\mathcal{F}_{ij} = -\left\langle \frac{\partial^{2}\mathcal{L}}{\partial\lambda_{i}\partial\lambda_{j}} \right\rangle.
\label{eq:fishermat}
\end{equation}
One of the key properties of the Fisher matrix is that the inverse of this matrix can be thought of as the best possible covariance matrix for a set of parameters based on the likelihood. As such the Fisher matrix provides a powerful tool to estimate the errors on cosmological parameters we can achieve for a given dataset. It is important to note that the inverse of Fisher matrix only gives the best possible \textit{statistical} errors on the parameters and does not inherently include any \textit{systematic} error budget. Any actual measurements of the velocity and density fields and subsequent cosmological constraints should also include a systematic error budget to account for inaccuracies in the modelling (such as the quoted accuracy of the RPT model at high-$k$) and possible measurement systematics. Hence we would expect any real cosmological constraints to be weaker than the forecasts presented here. We also explore some possible effects of ignoring modelling systematics on cosmological constraints in Section~\ref{sec:systematics}.

In our particular dataset we have a pair of Fourier-space observables $\delta_{g}(\bk)$ and $u(\bk)$ over a range of $k$-modes. First let us consider only a single $k$-mode. If we assume that this pair of observables are drawn from a multivariate Gaussian random distribution with mean vector $\boldsymbol{\xi}(\bk)$ and covariance matrix $\boldsymbol{C}(\bk)$, we can exploit the nature of the likelihood for a multivariate Gaussian and, substituting into Eq.~\ref{eq:fishermat}, obtain \citep{Vogeley1996,Tegmark1997a,Tegmark1997b}
\begin{equation}
\mathcal{F}_{ij}(\bk) = \frac{\partial{\boldsymbol{\xi}}^{T}}{\partial{\lambda_{i}}}\bC^{-1}\frac{\partial{\boldsymbol{\xi}}}{\partial{\lambda_{j}}} + 
		 \frac{1}{2}\,\mathrm{Tr}\left[\bC^{-1}\frac{\partial{\bC}}{\partial{\lambda_{i}}}\bC^{-1}\frac{\partial{\bC}}{\partial{\lambda_{j}}}\right].
\end{equation}
As the above expression is for a single $k$-mode our mean vector has two entries and we have a two-by-two covariance matrix. This covariance matrix is composed of the auto- and cross-correlations between the observables, which is related to the power spectra presented in Section~\ref{sec:power}. We will revisit this momentarily.

We can also write similar expressions for every mode we observe. Information is additive and so the full Fisher information in these two observables can be obtained by integrating over all the modes of interest within some volume $V$ \mbox{\citep{Mcdonald2009b}},
\begin{align}
F_{ij} &= V\int \frac{d^{3}k}{(2\pi)^{3}} \mathcal{F}_{ij}(k) \\
&= \frac{V}{2}\int \frac{d^{3}k}{(2\pi)^{3}}\,\mathrm{Tr}\left[\bC^{-1}\frac{\partial{\bC}}{\partial{\lambda_{i}}}\bC^{-1}\frac{\partial{\bC}}{\partial{\lambda_{j}}}\right]. \label{eq:fisher}
\end{align}
The integral is 3-dimensional and is taken over all components of the $\bk$ we are interested in. This can be simplified using spherical symmetry, as will be shown in Section~\ref{sec:fishercalc}. We arrive at the second equality because the first term in the Fisher matrix for each $k$-mode vanishes as $\langle \delta_{g}(\bk) \rangle = \langle u(\bk) \rangle = 0$. That is to say that the expected mean overdensity and peculiar velocity over the volume is expected to go to zero, and so the mean of our multivariate Gaussian is zero. However, the second term containing the covariance matrix for the two tracers remains.

\subsection{Covariance matrix and measurement noise} \label{sec:fishernoise}

The covariance matrix encapsulates both the cosmological information within the two fields via their auto- and cross-correlations, and, being the correlations of observed properties, the noise properties inherent in both fields which we have neglected up till now. In this section we will formulate the covariance matrix required to estimate the Fisher information matrix and introduce noise terms for the density and peculiar velocity measurements.

First, taking the covariance matrix as the correlations between our observables and combining this with the power spectra models developed in Section~\ref{sec:power} it should be apparent we can write
\begin{equation}
\boldsymbol{\bC}(\bk) = 
\begin{bmatrix}
P^{AA}_{gg}(\bk) & P^{AA}_{ug}(\bk) \\
P^{AA}_{ug}(\bk) & P^{AA}_{uu}(\bk)
\end{bmatrix}.
\label{eq:cov}
\end{equation}

However, both of our observables have noise terms which also contribute to the covariance matrix. Firstly, the positions of galaxies with respect to the underlying dark matter density perturbation can be seen as a stochastic process, and so the galaxy overdensity has some stochastic noise associated with it. This can be added to the galaxy overdensity as $\delta_{g}(\bk) \rightarrow \delta_{g}(\bk) + \epsilon$ where $\epsilon$ denotes the stochastic noise with properties $\langle \epsilon \rangle = 0$ and variance $\sigma^{2}_{\epsilon}$.

When measuring the galaxy-galaxy power spectrum using a measurement of the galaxy overdensity, this stochastic noise then enters as $P_{gg}(\bk) \rightarrow P_{gg}(\bk) + \sigma^{2}_{\epsilon}$, where the additional noise term is know as `shot-noise'. Assuming that the distribution of galaxies is a Poisson-point process, we can write the shot-noise in terms of the number density of galaxies $\bar{n}_{g}(\br)$ as $\sigma^{2}_{\epsilon} = 1/\bar{n}_{g}(\br)$.

For the velocity field there is also a contribution from the galaxy shot noise, $1/\bar{n}_{u}(\br)$. We let the number densities of density and velocity tracers be independent, as denoted by the different subscripts. Typically only a subset of a sample of galaxies with measured redshifts will have measured peculiar velocities due to the increased signal-to-noise required. There is also an additional error on the measurements of the peculiar velocities themselves $\sigma_{obs}(\br)$. This arises from the scatter in the astrophysical relations used to measure the peculiar velocity. We could add a similar error to the redshift measured for the density sample, however for a sample drawn from spectroscopic survey the redshift errors are typically negligible, whilst this is not true for measurements of the peculiar velocity. The total noise term for the peculiar velocity sample is then $\sigma^{2}_{obs}(\br)/n_{u}(\br)$ \citep{Burkey2004}.

In this study we assume that the number density of a given sample and the noise in the velocity measurements is roughly constant across the sky area of the survey (although, as will be shown in Section~\ref{sec:Fisher_2tracer}, variations in the sky coverage could be accounted for by computing separate Fisher matrices for the different sky areas and adding them). Instead we assume that the number density is only a function of the comoving distance between the observer and the object, $r$. In particular we assume that the noise in the velocity measurements consists of a constant fractional error, $\alpha$, multiplied by the distance to the object. This is generally a good assumption based on measurements for current surveys such as 6dFGSv and 2MTF \citep{Hong2014, Johnson2014}. To account for the velocity dispersion caused by the random non-linear motions of the galaxies within their host halos we also add a constant term $\sigma_{obs,rand}$. The total error in the peculiar velocity measurements is then
\begin{equation}
\sigma^{2}_{obs}(r) = (\alpha H_{0}r)^{2} + \sigma^{2}_{obs,rand}.
\label{eq:err}
\end{equation}
We adopt a constant value of $\sigma_{obs,rand}=300\,\mathrm{kms^{-1}}$, but the fractional error differs depending on the method used to obtain the peculiar velocities and will be given in Section~\ref{sec:data}. The additional presence of a zero-point offset, as already detailed in Section~\ref{sec:zp}, can also be incorporated into $\sigma_{u}$. However it is important to note that the origin and form of these two shot-noise-like components is very different, with the two original, linear, components of $\sigma_{obs}$ arising from statistical errors, whilst the zero-point offset is a systematic error, logarithmic in nature.

Combining the intrinsic correlations and noise for the two fields results in
\begin{equation}
\boldsymbol{\bC}(\br,\bk) = 
\begin{bmatrix}
P^{AA}_{gg}(\bk) + \frac{1}{\bar{n}^{A}_{g}(\br)} & P^{AA}_{ug}(\bk) \\
P^{AA}_{ug}(\bk) & P^{AA}_{uu}(\bk) + \frac{(\sigma^{A}_{obs}(\br))^{2}}{\bar{n}^{A}_{u}(\br)}
\end{bmatrix}.
\label{eq:cov}
\end{equation}
The off-diagonal terms in the covariance matrix do not contain any noise terms. Generally the noise between two fields or two surveys may be correlated, but in the case of stochastic, Poissonian shot-noise there is no correlation. Additionally, we expect the errors in the peculiar velocities, which are the main source of noise in the velocity power spectrum, to be uncorrelated with the noise in the density field, or between multiple surveys.

\subsection{Calculating the Fisher matrix} \label{sec:fishercalc}

When we include the intrinsic correlations and observational noise the nature of the covariance matrix is such that it is both a function of spatial coordinates $\br$ and Fourier modes $\bk$. However, under the `classical approximation' the volume $V$ in Eq.~\ref{eq:fisher} can be replaced with the integral $\int d^{3}x$ \citep{Hamilton1997,Abramo2012,Koda2014}. This approximation can be used so long as the wavelengths of the modes of interest in the power spectra are much smaller than the scale over which the noise varies (typically the size of the survey). In this case
\begin{equation}
F_{ij} = \frac{1}{2}\int \frac{d^{3}xd^{3}k}{(2\pi)^{3}}\,\mathrm{Tr}\left[\bC^{-1}\frac{\partial{\bC}}{\partial{\lambda_{i}}}\bC^{-1}\frac{\partial{\bC}}{\partial{\lambda_{j}}}\right]
\end{equation}

As we have assumed that the number density and velocity error only vary radially, we can make use of spherical symmetry to simplify the integrals over the $\br$- and $\bk$-vectors,
\begin{multline}
F_{ij} = \frac{\Omega_{sky}}{4\pi^{2}} \int^{r_{max}}_{0} r^{2} dr \int^{k_{max}}_{k_{min}} k^{2} dk \int^{1}_{0} d\mu \,\\
\mathrm{Tr}\left[\bC^{-1}(r,k,\mu)\frac{\partial{\bC}(r,k,\mu)}{\partial{\lambda_{i}}}\bC^{-1}(r,k,\mu)\frac{\partial{\bC}(r,k,\mu)}{\partial{\lambda_{j}}}\right].
\label{eq:Fisherrad}
\end{multline}
We have reduced the three-dimensional real-space integral to a sky area measured in steradians, $\Omega_{sky}$, multiplied by the integral along the line of sight $r$ and the three-dimensional $k$-space integral to an integral over the length of the $\bk$-vector along the line-of-sight and the cosine of the angle between the line-of-sight and this vector, $\mu$. This is the same $\mu$ as was introduced in Section~\ref{sec:power}, and means that now our covariance matrix can be written purely in terms of the models introduced in that section.

The $k$ integral is typically taken over $k\in[k_{min},k_{max}]$ where for this study we take separate values of $k_{min}$ for each dataset and for the density and peculiar velocity measurements. For the density field $k_{min}=2\pi/L_{max}$ where $L_{max}$ is roughly the largest separation between two galaxies in the sample, whilst for the velocity field we assume $k_{min}=0$ as the velocity field still encodes information far beyond the boundaries of the survey. In practice these two different limits are imposed by removing elements from the covariance matrix where there is expected to be no contribution from the power spectrum, i.e., (using the form in Eq.~\ref{eq:cov}) $C_{11} = 0$ for $k<2\pi/L_{max}$.\\

\subsection{Fisher matrix for multiple tracers and surveys} \label{sec:Fisher_2tracer}

The final step is to look at how to calculate the Fisher matrix for multiple, partially- or fully-overlapping surveys, each of which contains measurements of galaxy redshifts and peculiar velocities. For two independent surveys of the same density and velocity fields, we now have four observables, two measurements of the density field and two of the line-of-sight peculiar velocity field. Hence, for each $k$-mode we have a data-vector containing 4 elements and a 16 element covariance. Following the same steps used throughout this section and decomposing the three-dimensional $\br$ and $\bk$ vectors as per Eq.~\ref{eq:Fisherrad} we find,

\begin{widetext}
\begin{equation}
\boldsymbol{\bC}(r,k,\mu) = 
\begin{bmatrix}
P^{AA}_{gg}(k,\mu) + \frac{1}{\bar{n}^{A}_{g}(r)} & P^{AA}_{ug}(k,\mu) & P^{AB}_{gg}(k,\mu)& P^{AB}_{gu}(k,\mu) \\
P^{AA}_{gu}(k,\mu) & P^{AA}_{uu}(k,\mu) + \frac{(\sigma^{A}_{obs}(r))^{2}}{\bar{n}^{A}_{u}(r)} & P^{AB}_{ug}(k,\mu) & P^{AB}_{uu}(k,\mu) \\
P^{BA}_{gg}(k,\mu) & P^{BA}_{gu}(k,\mu) & P^{BB}_{gg}(k,\mu) + \frac{1}{\bar{n}^{B}_{g}(r)} & P^{BB}_{gu}(k,\mu) \\
P^{BA}_{ug}(k,\mu) & P^{BA}_{uu}(k,\mu) & P^{BB}_{ug}(k,\mu) & P^{BB}_{uu}(k,\mu) + \frac{(\sigma^{B}_{obs}(r))^{2}}{\bar{n}^{B}_{u}(r)}
\end{bmatrix}
\label{eq:covmulti}
\end{equation}
\end{widetext}
Each element of this covariance matrix contains a power spectrum that we have previously presented in Section~\ref{sec:power}, and results from the auto- and cross-correlation of the density and velocity fields measurements from two distinct surveys. The noise terms on the diagonal of the covariance matrix differ for each survey \textit{and} for galaxies used to measure the  density and velocity fields. Cross-correlating two surveys eradicates the noise terms (hence improving the constraining power) as the noise is assumed to be uncorrelated between the two surveys and so there are no noise terms included in any of the off-diagonal elements of the covariance matrix.

To actually calculate the Fisher matrix with this covariance matrix we must modify our method sightly. For any two surveys there is no guarantee that they will overlap completely in both the angular and radial directions, for instance one survey could focus on the full-sky whilst the second is only in the southern hemisphere, or the redshift range of one survey could be deeper than that of the other. In order to evaluate the Fisher matrix it is necessary to split Eq.~\ref{eq:Fisherrad} into multiple parts. This is a perfectly viable option as information is additive and this  approach is valid under the classical approximation, the derivation of which can also simply be thought of as a sum of sub-Fisher matrices. Hence, this approach remains applicable under the condition that the modes of interest remain smaller than each sub volume considered. For the combination of surveys considered in this paper, the sub-volumes still remain large enough for this approximation to hold.

For two surveys the first split is over the angular coordinates, where the full Fisher matrix becomes the sum of the sub-Fisher matrices for the two non-overlapping sky areas and the overlap region.
\begin{equation}
F_{ij} = F_{ij}(\Omega_{sky,A}) + F_{ij}(\Omega_{sky,B}) + F_{ij}(\Omega_{sky,AB})
\end{equation}
Each of $F_{ij}(\Omega_{sky,A})$, $F_{ij}(\Omega_{sky,B})$ and $F_{ij}(\Omega_{sky,AB})$ is its own Fisher matrix, calculated using Eq.~\ref{eq:Fisherrad}, but with the corresponding covariance matrix elements removed. $F_{ij}(\Omega_{sky,A})$ is the Fisher matrix for survey `A' on its own, calculated using only those parts of the covariance matrix that \textit{do not} depend on survey `B' and taking as the sky-area \textit{only} the area that survey `A' covers but survey `B' does not. It should be apparent that the covariance matrix in this case will reduce to Eq.~\ref{eq:cov} and the Fisher matrix is exactly that that would be calculated if we had never even considered a second survey, \textit{except} for the fact we use a different sky-area.

We do the same for the second survey `B' to calculate $F_{ij}(\Omega_{sky,B})$. To calculate $F_{ij}(\Omega_{sky,AB})$ we use the full 16-element covariance matrix from Eq.~\ref{eq:covmulti} and take as the sky-area \textit{only} the overlapping area for both surveys.

As an example, take a full-sky survey `A' and a hemispherical survey `B' with the same redshift range. The full Fisher matrix is calculated as the sum of the sub-Fisher matrix for the northern $2\pi$ steradians of the sky with the covariance matrix given by Eq.~\ref{eq:cov}, plus the sub-Fisher matrix for the $2\pi$ southern steradians of the sky with all terms in Eq.~\ref{eq:covmulti} retained. In this case the second sample B has no non-overlapping contribution, $F_{ij}(\Omega_{sky,B}) = 0$ . Under the formalism presented here a good sanity check is that each element of the sub-Fisher matrix for survey `A' will be exactly half that computed if we were to ignore the second survey completely. However for the full Fisher matrix each element will be larger that for just survey `A' due to the addition of information from survey `B' in the southern hemisphere.

To calculate the full Fisher matrix for real surveys, even after splitting the full Fisher matrix into sub-matrices for different regions on the sky, the overlapping area must still split up further as the two surveys may not overlap fully in redshift. Just like we did with the sky coverage the term $F_{ij}(\Omega_{AB})$ can be further divided into three integrals over the radial coordinates,
\begin{multline}
F_{ij}(\Omega_{AB}) = \frac{\Omega_{sky,AB}}{4\pi^{2}} \int^{k_{max}}_{k_{min}} k^{2} dk \int^{1}_{0} d\mu \\
\sum^{3}_{\ell=1} \biggl(\int^{r_{max,\ell}}_{r_{min,\ell}} r^{2} dr\, \mathrm{Tr}\left[\bC^{-1}\frac{\partial{\bC}}{\partial{\lambda_{i}}}\bC^{-1}\frac{\partial{\bC}}{\partial{\lambda_{j}}}\right]\biggl)
\end{multline}
i.e., when computing $F_{ij}(\Omega_{AB})$ we use this equation rather than Eq.~\ref{eq:Fisherrad}. The integration limits $r_{min,\ell}$ and $r_{max,\ell}$ are the minimum and maximum comoving distances for each of the radial patches: where survey `A' and `B' overlap, where survey `A' has data but `B' does not, and vice versa. In each case the corresponding covariance matrix is used. For example if survey `A' extends from $r=0-80\mpcoh$, whilt survey `B' is from $r=50-100\mpcoh$, then $r_{min} = [0, 50, 80]\mpcoh$ and $r_{max} = [50, 80, 100]\mpcoh$. For the first integral, with $r_{min}=0\mpcoh$ and $r_{max}=50\mpcoh$ we use the covariance matrix for \textit{only} survey `A', but unlike when we calculate $F_{ij}(\Omega_{A})$, we still use the sky-area of the overlapping region for the summation to make sense. For the second integral, $\ell=2$, we use the full covariance matrix and for the third we use only the covariance matrix for survey `B'. Overall, splitting both in sky coverage and redshift means that we only use the full covariance matrix for the combined surveys, Eq.~\ref{eq:covmulti}, when we are dealing with the part of the cosmological volume in which both surveys have data, as would be expected.

One final thing to note is that with more spectra comes a larger number of $k_{min}$ depending on the type of spectra. For all spectra involving the velocity field for either survey we still take $k_{min}=0$ whilst for the two galaxy-galaxy spectra we take different $k_{min}$ based on the $L_{max}$ of each sample and the $L_{max}$ between samples.

\section{Datasets} \label{sec:data}

\subsection{2MASS Tully-Fisher Survey} \label{sec:data2mtf}
The 2MASS Tully-Fisher (2MTF, \citealt{Masters2008,Hong2013,Masters2014,Hong2014}) survey is an all-sky survey of $\sim 2000$ nearby, bright spiral galaxies, with measured redshifts and `true' distances derived from fitting the Tully-Fisher relation to measured HI line widths.

HI measurements of the galaxies are recovered from archival data in the Cornell HI digital archive \citep{Springob2005}, supplemented by additional observations by the Green Bank Telescope (GBT; \citealt{Masters2014}), the Parkes radio telescope \citep{Hong2013} and the ALFALFA survey \citep{Haynes2011}. The conversion of these measurements into estimates of the logarithmic distance ratio for each galaxy, that is the ratio between the `true' distance and the distance inferred from the galaxy's redshift, are presented in \cite{Hong2013,Hong2014}, alongside corrections for Malmquist bias.

The final 2MTF sample covers the full-sky except for galactic latitudes $|b|<5^{\circ}$, where galactic dust prevents accurate observations. For these forecasts we hence assume a sky-area of $3.65\pi$ steradians. It should be noted that the nature of the combined HI observations from the GBT, Parkes and ALFALFA data results in an inhomogeneous sky coverage for the 2MTF, with fewer galaxies with $\delta<-40^{\circ}$ than would be expected. The number density below $\delta=-40^{\circ}$ is approximately a factor of 2 lower than that above this declination. Bulk flow measurements using the 2MTF data account for this using a weighting scheme, as would future cosmological measurements. However, for this paper we do not treat the two separate sky areas separately, as we expect this inhomogeneity will have little impact on the forecasts.

The number density of the 2MTF sample is shown in Figure~\ref{fig:nz} alongside the other samples used in the study. For the forecasts in this paper we assume a fractional velocity error of $\alpha=0.22$ \citep{Hong2014} and $k_{min}=0.032\hompc$ for the forecasts using the density field.

A key parameter in the Fisher matrix forecasts involving the density field is the galaxy bias for a particular sample. Several past studies have looked at the galaxy bias of HI selected samples, including data from the same surveys that are used to form the 2MTF sample. Typical values for the bias of neutral hydrogen with respect to the underlying dark matter measured using simulations and observations are found be be $\sim 0.7-1.0$ with an uncertainty of 0.2 \citep{Basilakos2007, Martin2012, Dave2013, Hoppmann2015, Padmanabhan2015}. However, though the 2MTF galaxies are chosen to be gas-rich, the sample itself is selected from IR photometry, and so would be typically expected to have a higher bias than a fully HI selected sample. Looking at forecasts for the 2MTF survey using different values for the bias between $0.7$ and $1.0$, we find all of our constraints to be insensitive to the exact value that we use. The velocity field forecasts are independent of the bias, and for the combined velocity and density forecasts for the 2MTF sample alone and in combination with other surveys, the majority of the information and improvement on $f\sigma_{8}$ still comes from the peculiar velocity measurements. As such we adopt a value of $b=1.0$ for all our quoted forecasts.

\subsection{6-degree Field Galaxy Survey}
The 6-degree Field Galaxy Survey (6dFGS, \citealt{Jones2004,Jones2005,Jones2009}) is a combined galaxy redshift and peculiar velocity survey of early-type galaxies within $z\leq0.15$, which covers the full southern sky with the exception of the region about the galactic plane with $|b|<10^{\circ}$. The velocity subsample consists of galaxies with $z\leq0.05$, with peculiar velocities derived using the Fundamental plane relation, calibrated in \cite{Magoulas2012}. The full Fundamental plane catalogue is presented in \cite{Campbell2014} and subsequent measurements of the logarithmic distance ratios for the galaxies are given in \cite{Springob2014}. 

\begin{figure}
\centering
\includegraphics[width=0.5\textwidth]{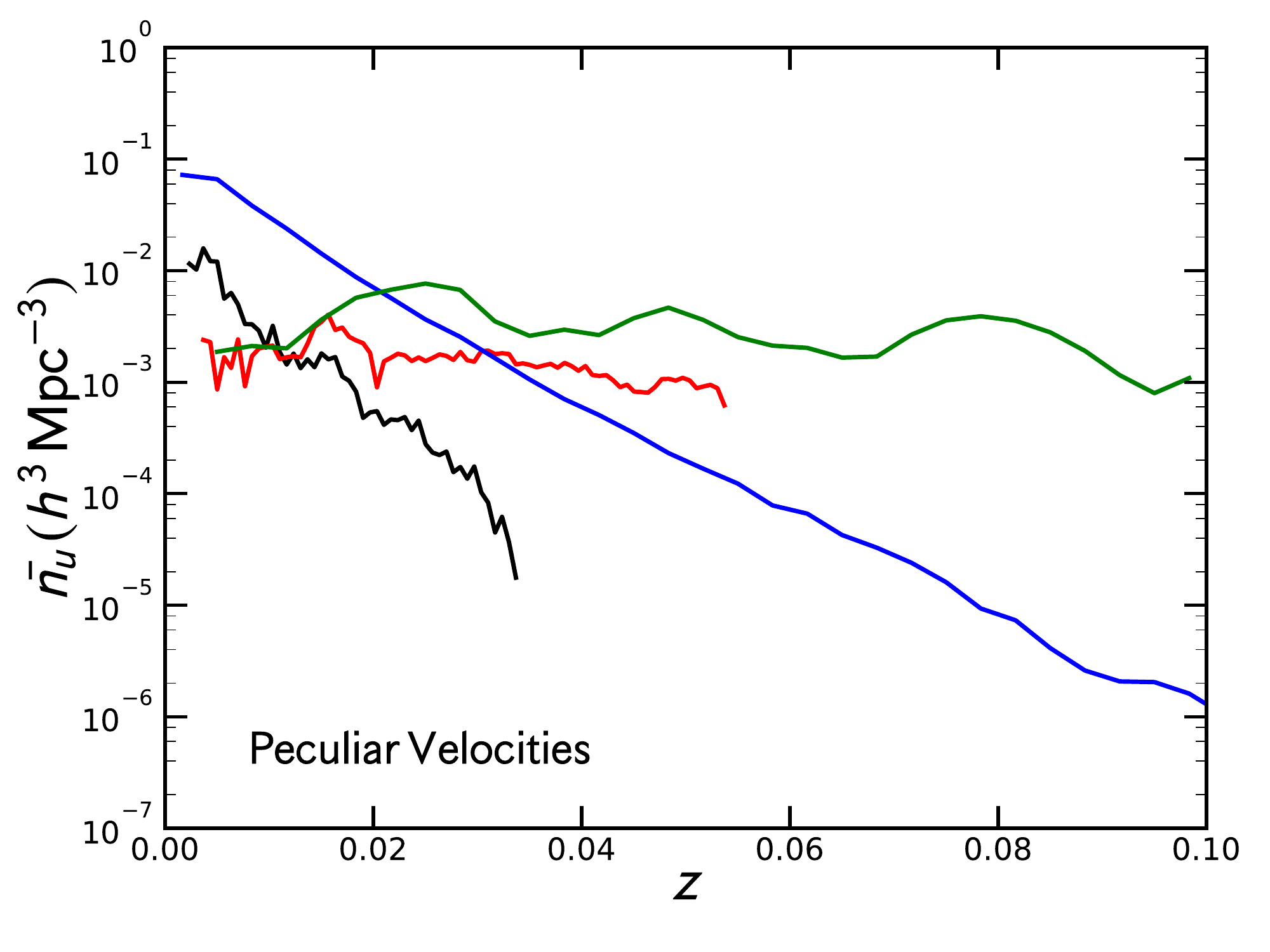} \\
\includegraphics[width=0.5\textwidth]{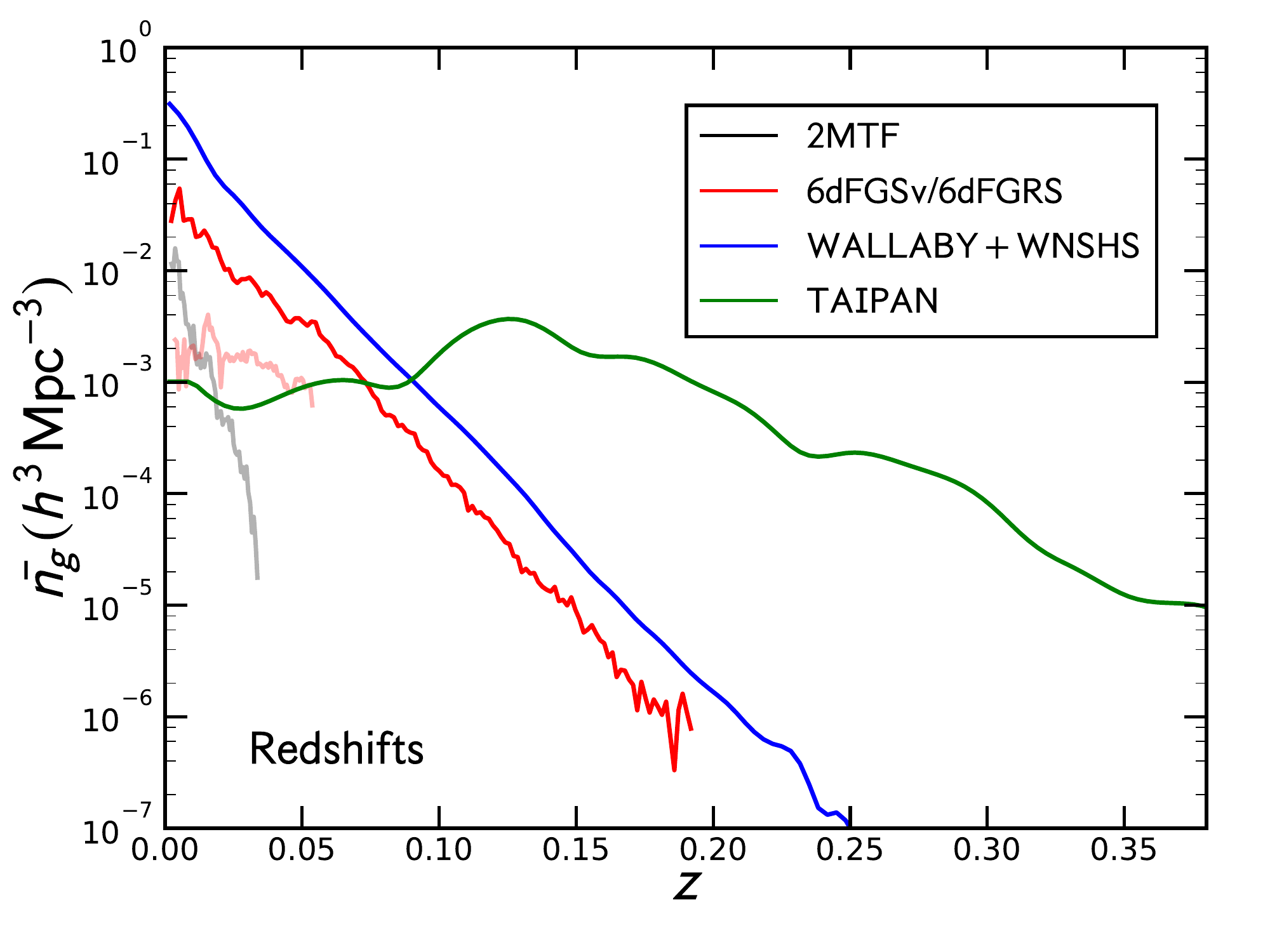}
  \caption{The number density as a function of redshift for all the peculiar velocity (top) and redshift (bottom) surveys considered in this paper. The faded lines in the lower panel are the number densities of the 2MTF and 6dFGSv peculiar velocity samples replotted for comparison with the other redshift surveys, as for these surveys we also consider the case where $\bar{n}_{g}=\bar{n}_{u}$.}
  \label{fig:nz}
\end{figure}

Both the redshift sample, containing $\sim110,000$ galaxies, and the peculiar velocity subsample of $\sim8800$ galaxies will be used within this study, to forecast constraints on measuring the velocity and density fields from the velocity subsample alone, or through combination with the full redshift survey. The number density of these two samples is shown in Figure~\ref{fig:nz}. Fisher matrix forecasts for the density field only and for the combination of density and velocity fields measured from the velocity sample have been presented in \cite{Beutler2012} and \cite{Koda2014} respectively. In this study we go one further and look at the combined constraints from the velocity subsample and full redshift survey combined, and from combinations of 6dFGS galaxies with other surveys. For consistency with the previous works, we adopt the same bias parameter $b=1.4$ and sky coverage $\Omega_{sky} = 1.65\pi$ steradians. We use a fractional distance error of $\alpha=0.26$ as quoted in \cite{Johnson2014} and $k_{min}=0.02\hompc (0.008\hompc)$ for the forecasts using the density field from the velocity subsample and full redshift survey respectively.

\subsection{The TAIPAN survey} \label{sec:datataipan}
The TAIPAN survey (da Cunha et. al., in prep.) is the spiritual successor to the 6dGFS, aiming to obtain spectra for over 1,000,000 low redshift galaxies over the full southern sky. Of these 1,000,000 spectra a large number are also expected to have sufficient signal-to-noise to enable peculiar velocity measurements via the fundamental plane relation. Commissioning of the survey is planned to begin in the first half of 2017 on the refurbished UK-Schmidt telescope. The key science goal of the four-year galaxy survey is to obtain a 1\% precision measurement of the Hubble parameter. However, the galaxy redshift and peculiar velocity samples will also have significant impact on measurements of the growth rate of structure and galaxy properties in the low redshift universe.

Forecasts of the constraints on the growth rate of structure from the TAIPAN survey have been produced by both \cite{Beutler2012} for the redshift survey alone, and by \cite{Koda2014} for the combined redshift and peculiar velocity sample. Here we expand these forecasts to include updated estimates of the number density of tracers from the TAIPAN survey, look at possible systematics that could bias constraints and identify whether we can expect any constraints on additional cosmological parameters using this dataset. Additionally, we also look at the benefits of combining the TAIPAN and WALLABY+WNSHS surveys (the latter of which will be detailed in the next section). Although \cite{Beutler2012} found little benefit in combining the two redshift surveys using the multi-tracer methodology \citep{Mcdonald2009b}, we investigate any gains obtained when combining the full datasets, including the redshift and peculiar velocity subsamples of both surveys, and overlapping and non-overlapping regions.

How the estimated number densities of the final TAIPAN redshift and peculiar velocity samples were obtained will be presented in a forthcoming paper (da Cunha et. al., in prep.). To summarise, the estimates are based on a smaller number of objects with known redshifts and photometry obtained from cross-matching data from the Sloan Digital Sky Survey Data Release 10 (SDSS-DR10; \citealt{Ahn2014}) and Two Micron All Sky Survey (2MASS; \citealt{Skrutskie2006}). Redshift targets were selected using an $i$-band magnitude limit of 17.5, and $g-i>1.4$ colour cut. Peculiar velocity targets were selected using a J-band magnitude limit of 15.4, a redshift limit of $z<0.1$, cuts in the H$\alpha$, D4000 and H$\delta$ spectral features to preference early type galaxies, and signal-to-noise and velocity dispersion limits of 6.5 and $60\,\mathrm{kms^{-1}}$. These choices result in an estimated $\sim100,000$ objects in the southern hemisphere for which it is feasible for the TAIPAN survey to measure peculiar velocities over the course of the survey.

For our forecasts we assume a galaxy bias of $b=1.2$, based on preliminary clustering measurements of a subset of prospective targets, and a sky area of $\Omega=1.65\pi$ steradians. Although we assume the same sky area as the 6dFGS, it is feasible that the TAIPAN survey will push to somewhat higher declinations, increasing the sky area beyond that assumed here. This would increase the cosmological volume probed and hence makes these estimates conservative. The effective redshift of the TAIPAN survey is expected to be $z\approx0.19$, though there will still exist some small number of galaxies up to $z=0.4$. Due to this increase in redshift depth compared to the 6dFGS we use a smaller value of $k_{min}=0.003\hompc$. We use a fractional error of $\alpha=0.2$ as in \cite{Koda2014}, reflecting the belief that we will have slightly better knowledge of the Fundamental Plane relation (in particular how the scatter is affected by morphology) than was available with the 6dFGS peculiar velocity sample. When looking at the effects of scale-dependent and velocity bias we use values of $b_{\zeta}=7.9h^{-2}\,{\rm Mpc}^{2}$ and $R_{v}=2.6\mpcoh$. These normalisations are calculated using the method detailed in Section~\ref{sec:velbias} for halos with $M_{halo}=10^{13}h^{-1}M_{\odot}$, which we expect to be typical for the halos in which the TAIPAN galaxies reside.

\subsection{WALLABY+WNSHS}
The Widefield ASKAP L-band Legacy All-sky Blind Survey (WALLABY; \citealt{Johnston2008}) is a planned HI survey using the Australian SKA Pathfinder (ASKAP), covering three quarters of the full-sky up to $z=0.25$. This is complemented by the planned Westerbork Northern Sky HI survey (WNSHS), which is expected to eventually cover the remaining quarter of the sky. Combining HI simulations with the expected instrumental throughput of these two surveys results in a predicted $\sim 800,000$ galaxies with redshift measurements determined using 21-cm line \citep{Duffy2012}. A significant fraction of these $\sim 40,000$ are also expected to have peculiar velocity measurements determined via the Tully-Fisher relation. 

As with the TAIPAN survey, in this paper we expand on forecasts that have already been produced by \cite{Beutler2012} and \cite{Koda2014}. Whilst at the lower limit for a HI selected survey (see Section~\ref{sec:data2mtf}), for consistency with their results we adopt a galaxy bias of $b=0.7$ \citep{Basilakos2007} and fractional distance uncertainty $\alpha=0.2$. Based on the maximum redshift of $z=0.25$ we use a value $k_{min}=0.0045\hompc$ for the forecasts utilising the density field information. 

From the simulated HI mass function of \cite{Duffy2012}, we expect the galaxies observed by the combined WALLABY and WHSNS surveys to reside in halos with typical mass $M_{halo}=10^{12}h^{-1}M_{\odot}$. As such we adopt lower values of $b_{\zeta}=1.4h^{-2}\,{\rm Mpc}^{2}$ and $R_{v}=1.3\mpcoh$ than those used for the TAIPAN dataset.

\begin{table*}
\caption{Predicted percentage uncertainties on the growth rate $f$, and non-linear RSD damping of the velocity power spectrum parameterised by $\sigma_{u}$ using \textit{only} the peculiar velocity measurements, i.e., using only the observable $u(\bk)$ and \textit{not} $\delta_{g}(\bk)$. The middle two columns show the percentage errors using $k_{max}=0.1\hompc$, whilst the last two show the percentage errors using $k_{max}=0.2\hompc$. For each survey or combination of surveys listed in the first column, we list the parameters included in that particular forecast in the second column, i.e., $f\sigma_{8}$ alone means we ignore the non-linear RSD, whilst $f\sigma_{8},\,\sigma_{u}$ means we marginalise over it. When combining two surveys we have two independent nuisance parameters and hence two values for the percentage error, presented in the same order as the survey names.}
\centering
\begin{tabular}{llcccc} \hline
\multicolumn{2}{c}{Velocity Field Only} & \multicolumn{2}{c}{$\quad k_{\mathrm{max}} = 0.1h\mathrm{Mpc}^{-1}$} & \multicolumn{2}{c}{$\quad k_{\mathrm{max}} = 0.2h\mathrm{Mpc}^{-1}$}\\
Survey  			& Parameters & $100\times\sigma{(f\sigma_{8})}\,/\,f\sigma_{8}$ & $100\times\sigma{(\sigma_{u})}\,/\,\sigma_{u}$ & $100\times\sigma{(f\sigma_{8})}\,/\,f\sigma_{8}$ & $100\times\sigma{(\sigma_{u})}\,/\,\sigma_{u}$ \\ \hline
2MTF    			& $f\sigma_{8}$ 				& 30.3 &   -   & 26.7 &   -  \\
	     		& $f\sigma_{8},\,\sigma_{u}$ 	& 51.2 & 179.0 & 35.1 & 50.7 \vspace{5pt} \\ 
6dFGSv   		& $f\sigma_{8}$ 				& 25.1 &   -   & 24.3 &   -  \\
	     		& $f\sigma_{8},\,\sigma_{u}$ 	& 39.2 & 170.5 & 32.6 & 95.9 \vspace{5pt} \\ 
2MTF +			& $f\sigma_{8}$ 				& 20.5 &   -   & 18.8 &   -  \\
6dFGSv			& $f\sigma_{8},\,\sigma_{u}$ 	& 33.2 & 134.7, 145.2 & 24.7 & 43.7, 73.5 \vspace{5pt} \\ 
TAIPAN   		& $f\sigma_{8}$ 				& 10.2 &   -  &  9.9 &   -  \\
	     		& $f\sigma_{8},\,\sigma_{u}$ 	& 16.0 & 68.5 & 13.1 & 36.0 \vspace{5pt} \\
WALLABY + WNSHS 	& $f\sigma_{8}$ 				& 13.0 &   -  & 10.7 &   -  \\
		    		& $f\sigma_{8},\,\sigma_{u}$ 	& 22.4 & 73.8 & 14.2 & 15.1 \vspace{5pt} \\ 
TAIPAN +	 		& $f\sigma_{8}$ 				&  8.5 &   -  &  7.5 &   -  \\
WALLABY + WNSHS	& $f\sigma_{8},\,\sigma_{u}$ 	& 13.7 & 58.6, 54.2 &  9.8 & 27.2, 13.2 \vspace{5pt} \\  
\end{tabular}
\label{tab:vel}
\end{table*}

\begin{figure*}
\centering
\subfloat{\includegraphics[width=0.51\textwidth]{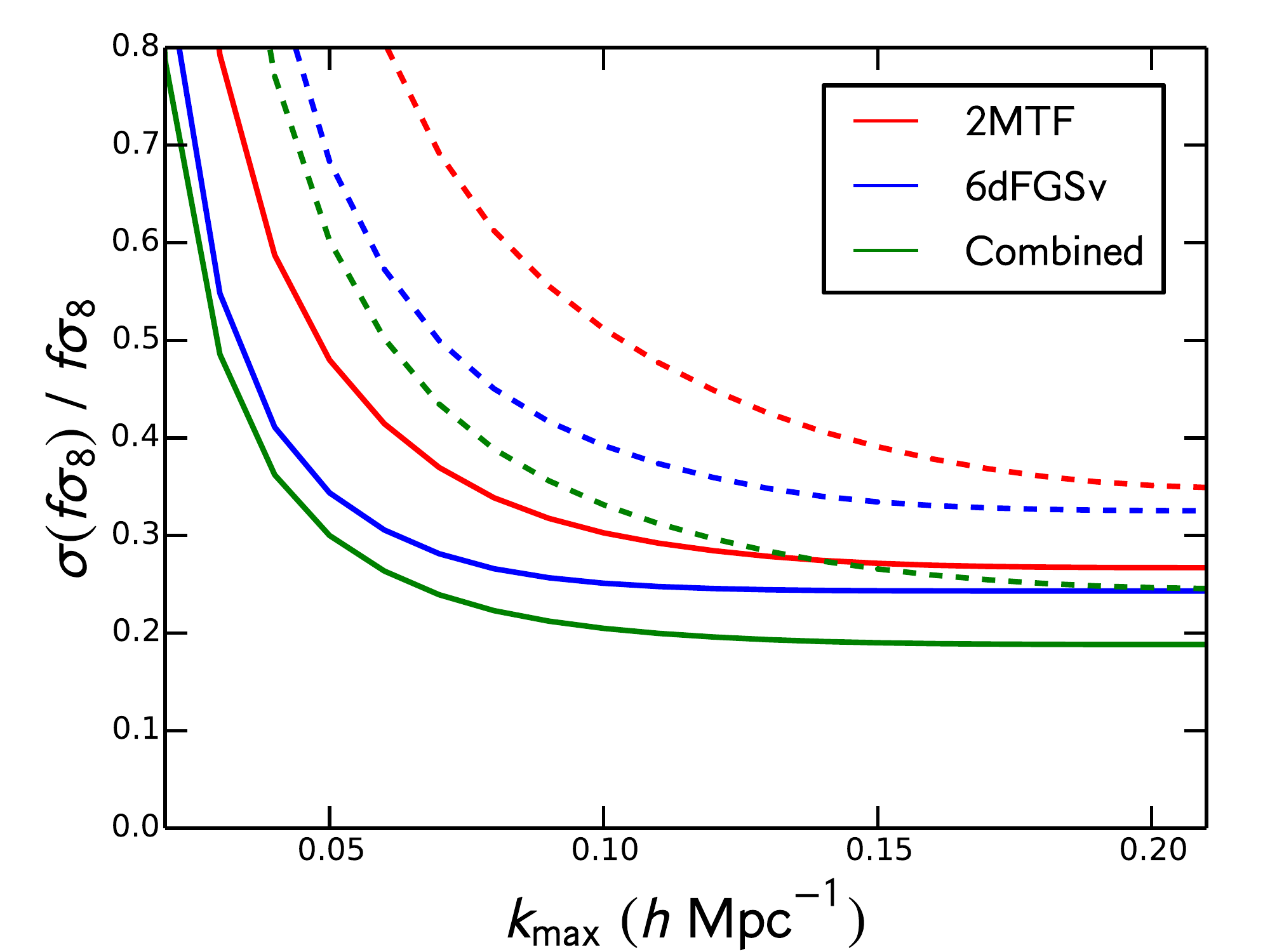}}
\subfloat{\includegraphics[width=0.51\textwidth]{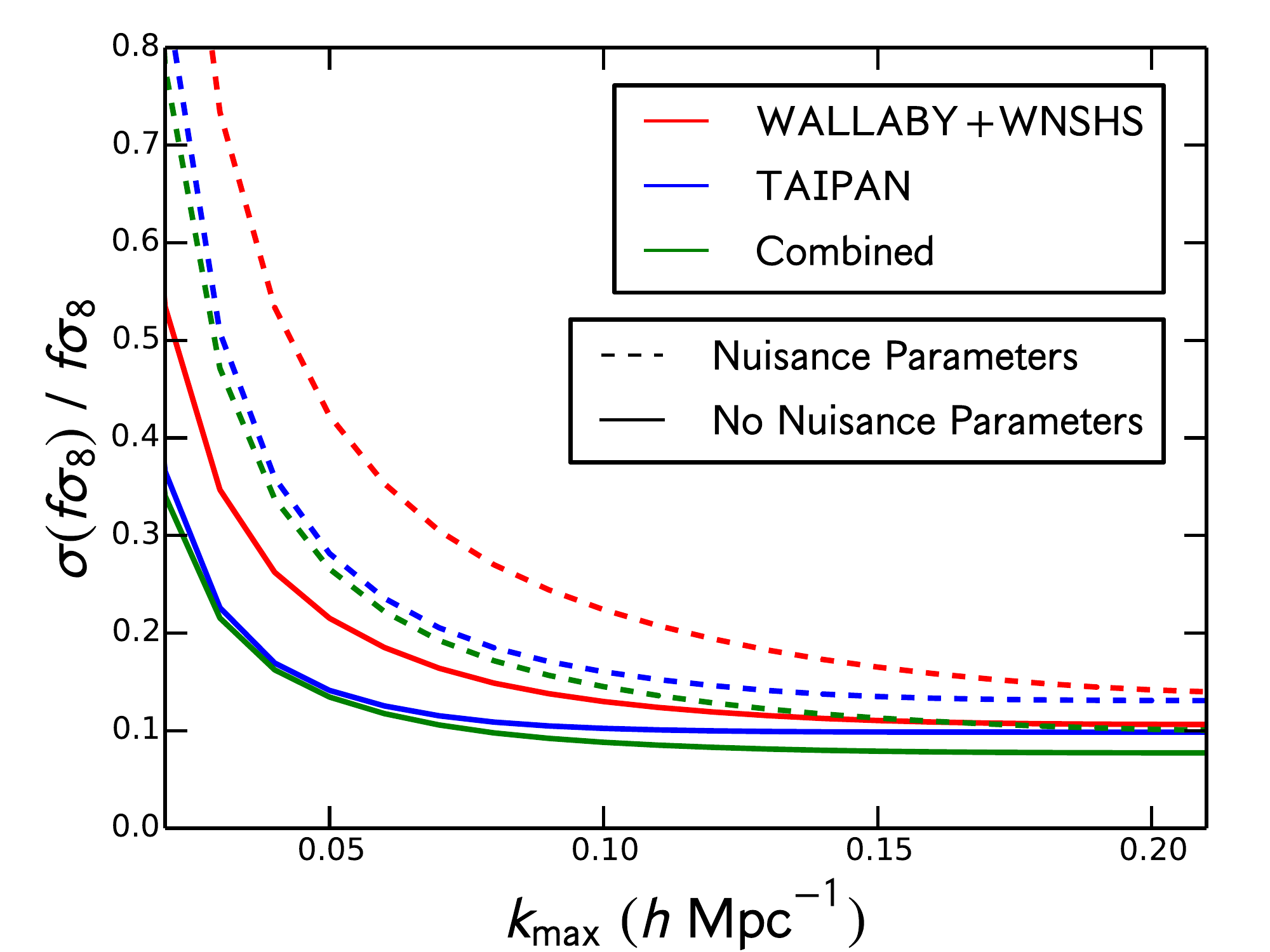}}
  \caption{$1\sigma$ percentage errors on $f\sigma_{8}$ for \textit{only} the peculiar velocities of the 2MTF, 6dFGSv, WALLABY+WNSHS and TAIPAN surveys as a function of the maximum $k$ used in the forecasts. These forecasts use only the information in the line-of-sight peculiar velocities, the observable $u(\bk)$, and hence correspond to the predictions in Table~\ref{tab:vel} when $k_{max} = 0.1\hompc$ and $0.2\hompc$. Higher $k_{max}$ signifies the inclusion of more non-linear information. Solid lines show the forecasts assuming perfect knowledge of the nuisance parameter, $\sigma_{u}$, whilst the dashed lines are when this is marginalised over. In the case of two surveys combined, which are shown in green with 2MTF+6dFGSv in the left panel and TAIPAN+WALLABY+WNSHS in the right, separate nuisance parameters for each survey are used.}
  \label{fig:vel}
\end{figure*}

\section{Fisher matrix results} \label{sec:results}

In this section we investigate the Fisher matrix forecasts for the growth rate constraints from the surveys and datasets described in Section~\ref{sec:data}. We look at the constraints from the velocity field alone for each survey and when the two current and next generation datasets are combined (2MTF+6dFGSv and TAIPAN+WALLABY+WNSHS). We then use our 4 tracer Fisher matrix formalism to look at the constraints from combining the velocity and density fields for each survey alone and then in combination. In particular we look at the combinations of the 2MTF and 6dFGSv datasets, 2MTF and the full 6dFGS sample, consisting of both the redshift survey and peculiar velocity subsample, and the combination of TAIPAN and WALLABY+WNSHS.

For a given survey or set of surveys we obtain the predicted errors on the cosmological parameters by evaluating Eq.~\ref{eq:Fisherrad} (or the equivalent expressions if we need to split into radial and angular patches for combined surveys) with the covariance matrix calculated using the models from Section~\ref{sec:power} and fiducial cosmological and noise parameters from Sections~\ref{sec:params} and~\ref{sec:data}. In all cases, except when looking at the $\gamma$ parameter (Section~\ref{sec:gammaconstraints}) we compute the derivatives of the covariance matrix analytically, putting the corresponding parameter values directly into these expressions. The Fisher matrix is then inverted to obtain the covariance matrix for the cosmological parameters we are including and the errors on those parameters are extracted from this parameter covariance matrix. 

When we include the nuisance parameters $r_{g}$, $\sigma_{g}$ and $\sigma_{u}$ we assume no prior information, as such these forecasts can be seen as conservative. For the cases where we look at the constraints ignoring nuisance parameters the nature of the Fisher matrix is such that we can simply remove the rows and columns from the matrix corresponding to those parameters before inverting to get the parameter covariance matrix. The resulting parameter covariance matrix is as if we had not marginalised over these parameters, or equivalently, assumed perfect prior knowledge of their behaviour.

\subsection{Measurements from peculiar velocities alone}

The forecasts for the growth rate for $k_{max}=0.1\hompc$ and $0.2\hompc$ (the upper integration limit in Eq.~\ref{eq:Fisherrad}) using only the peculiar velocity samples for each survey, and with and without nuisance parameters, is shown in Table~\ref{tab:vel}. In this case we are only using information in the observable $u(\bk)$ and \textit{not} in $\delta_{g}(\bk)$. Hence, the covariance matrix used for the calculation of the Fisher matrix only has one element for a single survey and four for two surveys combined.

Using a higher value of $k_{max}$ in the Fisher matrix calculation means we are including more non-linear information, and so should improve the constraints on the growth rate, but at the cost of increased sensitivity to the measurement error, $\sigma_{obs}$, and to the non-linear effects of RSD (i.e., increased need to marginalise over $\sigma_{g}$ and $\sigma_{u}$). Because the measurement error acts as a shot-noise like component to the velocity power spectrum independent of $k$, it becomes more and more dominant on non-linear scales as the intrinsic velocity power spectrum gets smaller. This in turn means that there is some scale at which including more $k$-modes has little effect on the growth rate constraints. A more complete picture is given in Fig~\ref{fig:vel} where we plot the constraints on the growth rate with and without nuisance parameters as a function of $k_{max}$.

We find that our forecasts for the 6dFGSv and WALLABY+WNSHS surveys are in generally good agreement with those reported in \cite{Koda2014}. Our TAIPAN constraints are tighter than those of \cite{Koda2014} due to significant differences in the number densities used in their study and in ours. In all cases we find a significant degradation of the results due to the effect of having to marginalise over the unknown RSD damping parameter $\sigma_{u}$. We would expect that some prior information on this parameter may be available through simulations that reproduce the peculiar velocity samples, which could improve the constraints.

Although the number density is on average lower, we find that the 2MTF peculiar velocities can provide an interesting, independent measurement of the growth rate. The 2MTF survey has a significantly higher density of galaxies at the lowest redshifts where peculiar velocity measurements are more accurate and the overall lower number density is partially compensated for by the lower percentage distance error (our parameter $\alpha$ in Eq.~\ref{eq:err}) obtained when using the Tully-Fisher relation rather than the Fundamental Plane relation, the former of which has slightly less intrinsic scatter. This means that though the amount of information on purely linear scales is less (comparing the $k_{max}=0.1\hompc$ forecasts), the gain when fitting to smaller scales is greater for the 2MTF survey and the constraints on the RSD damping are stronger, so the effect of marginalising over this on the growth rate measurements is less. This means that overall the constraints when including $\sigma_{u}$ for $k_{max}=0.2\hompc$ are remarkably similar.

We also find that combining the 2MTF and 6dFGSv datasets improves the constraints from the two surveys alone, purely in terms of statistical power, by $\approx 25\%$ for $k_{max}=0.2\hompc$, regardless of whether or not the RSD damping term is marginalised over. However, the actual gains are likely to be even better as combining with 2MTF will be able to reduce the effects of unknown systematics.

When looking at the redshift sample \cite{Beutler2012} found little improvement on the growth rate constraints when combining the WALLABY+WNSHS and TAIPAN surveys. However, we find that instead there does seem to be merit in combining the peculiar velocity samples of TAIPAN and WALLABY+WNSHS. This could be due in part to the now higher number density of TAIPAN improving the overlap effect between the two. In this study we have also combined the full sample, rather than just approximating the effective area of the overlap volume, which we expect to more accurately reflect the improvement. Overall, combining these two velocity samples improves the growth rate constraints by 15-30\% compared to their individual measurements.

\subsection{Measurements from peculiar velocities and redshifts}

\begin{table*}
\caption{Fisher matrix forecasts for the percentage uncertainties on cosmological parameters using information in both the velocity and density fields, i.e, including both the observables $\delta_{g}(\bk)$ and $u(\bk)$ in the Fisher matrix calculation. The format is similar to Table~\ref{tab:vel}, except we include the forecasts for $k_{max}=0.1\hompc$ as the upper block, and the the forecasts for $k_{max}=0.2\hompc$ as the lower block. For each survey and combination of surveys we look at the case with and without the nuisance parameters $r_{g}$, $\sigma_{g}$, $\sigma_{u}$, but always include the effects of galaxy bias in the form of $\beta$. For two surveys there are two distinct fiducial values for all of these parameters and hence two sets of predicted percentage errors. All the surveys we consider are the same as those in Table~\ref{tab:vel} except we also consider the additional case where we take just the density field information from the 6dFGSv sample and when we combine it with the remaining redshifts and density field information from the full 6dFGRS catalogue.}
\centering
\begin{tabular}{llccccc} \hline
\multicolumn{2}{c}{Combined Density and Velocity Fields} & \multicolumn{5}{c}{$100\times\sigma(\theta_{i})\,/\,\theta_{i}$} \\
Survey   & Parameters & $f\sigma_{8}$ & $\beta$ & $r_{g}$ & $\sigma_{u}$ & $\sigma_{g}$ \\ \hline
\multicolumn{7}{r}{$k_{max} = 0.1\hompc$} \vspace{3pt} \\
2MTF			    & $f\sigma_{8},\,\beta$ 									& 19.4 & 19.1 &   -  &   -  &   -  \\
	     		& $f\sigma_{8},\,\beta,\,r_{g},\,\sigma_{u},\,\sigma_{g}$	& 33.9 & 33.3 &  3.6 & 113.9 & 622.6 \vspace{5pt} \\
6dFGSv   		& $f\sigma_{8},\,\beta$ 								  	& 15.9 & 16.3 &   -  &   -  &   -  \\
	     		& $f\sigma_{8},\,\beta,\,r_{g},\,\sigma_{u},\,\sigma_{g}$	& 24.9 & 24.3 &  4.7 & 103.1 & 370.4 \vspace{5pt} \\
6dFGSv + 		& $f\sigma_{8},\,\beta$ 								  	& 11.2 & 12.3 &   -  &   -  &   -   \\
6dFGRS	 		& $f\sigma_{8},\,\beta,\,r_{g},\,\sigma_{u},\,\sigma_{g}$	& 16.7 & 17.2 &  1.8 & 83.9 & 143.0 \vspace{5pt} \\
2MTF +   		& $f\sigma_{8},\,\beta$ 									& 12.4 & 13.9, 12.5 &   -  &   -  &   -  \\
6dFGSv	     	& $f\sigma_{8},\,\beta,\,r_{g},\,\sigma_{u},\,\sigma_{g}$	& 20.0 & 21.5, 19.4 &  3.2, 3.1 & 80.9, 90.2 & 462.1, 30.9 \vspace{5pt} \\
2MTF +   		& $f\sigma_{8},\,\beta$ 									&  9.0 & 12.1,  9.8 &   -  &   -  &   -  \\
6dFGSv +	 6dFGRS	& $f\sigma_{8},\,\beta,\,r_{g},\,\sigma_{u},\,\sigma_{g}$	& 13.8 & 17.1, 14.0 &  2.7, 1.0 & 67.8, 77.1 & 358.1, 136.4 \vspace{5pt} \\
TAIPAN   		& $f\sigma_{8},\,\beta$ 								  	&  4.2 &  4.7 &   -  &   -  &   -  \\
	     		& $f\sigma_{8},\,\beta,\,r_{g},\,\sigma_{u},\,\sigma_{g}$	&  7.0 &  7.3 &  2.3 & 34.8 & 46.3 \vspace{5pt} \\ 
WALLABY + 		& $f\sigma_{8},\,\beta$ 									&  4.0 &  4.6 &   -  &   -  &   -  \\
WNSHS	  		& $f\sigma_{8},\,\beta,\,r_{g},\,\sigma_{u},\,\sigma_{g}$	&  6.3 &  6.5 &  0.3 & 25.5 & 86.1 \vspace{5pt} \\ 
TAIPAN + 		& $f\sigma_{8},\,\beta$ 								  	&  2.8 &  3.4,  3.2 &   -  &   -  &   -  \\
WALLABY + WNSHS 	& $f\sigma_{8},\,\beta,\,r_{g},\,\sigma_{u},\,\sigma_{g}$	&  4.6 &  4.7,  4.8 &  1.2, 0.3 & 28.7, 21.9 & 38.4, 62.0 \vspace{5pt} \\ 
\multicolumn{7}{r}{$k_{max} = 0.2\hompc$} \vspace{3pt} \\
2MTF			    & $f\sigma_{8},\,\beta$ 									& 14.8 & 16.5 &   -  &   -  &   -   \\
	     		& $f\sigma_{8},\,\beta,\,r_{g},\,\sigma_{u},\,\sigma_{g}$	& 20.8 & 21.2 &  3.5 & 27.4 & 92.6  \vspace{5pt} \\
6dFGSv   		& $f\sigma_{8},\,\beta$ 								  	& 12.8 & 14.0 &   -  &   -  &   -   \\
	     		& $f\sigma_{8},\,\beta,\,r_{g},\,\sigma_{u},\,\sigma_{g}$	& 17.6 & 17.9 &  4.7 & 32.8 & 45.7  \vspace{5pt} \\
6dFGSv + 		& $f\sigma_{8},\,\beta$ 								  	&  8.0 &  8.9 &   -  &   -  &   -   \\
6dFGRS	 		& $f\sigma_{8},\,\beta,\,r_{g},\,\sigma_{u},\,\sigma_{g}$	& 11.7 & 12.1 &  1.8 & 29.2 & 21.5  \vspace{5pt} \\
2MTF +   		& $f\sigma_{8},\,\beta$ 									&  9.7 & 11.4, 10.6 &   -  &   -  &   -   \\
6dFGSv	     	& $f\sigma_{8},\,\beta,\,r_{g},\,\sigma_{u},\,\sigma_{g}$	& 13.3 & 14.3, 13.5 &  3.2, 3.0 & 23.5, 30.3 & 71.6, 42.3 \vspace{5pt} \\
2MTF +   		& $f\sigma_{8},\,\beta$ 									&  6.8 &  8.6,  7.5 &   -  &   -  &   -   \\
6dFGSv +	 6dFGRS	& $f\sigma_{8},\,\beta,\,r_{g},\,\sigma_{u},\,\sigma_{g}$	&  9.7 & 11.2, 10.0 &  2.6, 1.0 & 22.0, 28.3 & 59.5, 20.0 \vspace{5pt} \\
TAIPAN   		& $f\sigma_{8},\,\beta$ 								  	&  2.3 &  2.6 &   -  &   -  &   -   \\
	     		& $f\sigma_{8},\,\beta,\,r_{g},\,\sigma_{u},\,\sigma_{g}$	&  4.1 &  4.2 &  2.3 & 12.1 &  6.8  \vspace{5pt} \\ 
WALLABY + 		& $f\sigma_{8},\,\beta$ 									&  2.7 &  3.3 &   -  &   -  &   -   \\
WNSHS	  		& $f\sigma_{8},\,\beta,\,r_{g},\,\sigma_{u},\,\sigma_{g}$	&  4.2 &  4.4 &  0.3 &  6.8 & 12.9  \vspace{5pt} \\ 
TAIPAN + 		& $f\sigma_{8},\,\beta$ 								  	& 1.8 &  2.2,  2.0 &   -  &   -  &   -  \\
WALLABY + WNSHS 	& $f\sigma_{8},\,\beta,\,r_{g},\,\sigma_{u},\,\sigma_{g}$	& 2.8 &  3.0, 3.1 &  1.1, 0.3 &  10.9, 6.4 &  5.7, 9.7 \vspace{5pt} \\ 

\end{tabular}
\label{tab:vel2}
\end{table*}

Modelling of the redshift space galaxy clustering has long been the preferred method of determining the growth rate of structure. Unlike the peculiar velocity samples, the only requirement is for enough signal to noise in each galaxy spectra to reliably measure a redshift,  which allows us to cover large cosmological volumes with high number density. However, even on linear scales, the biased way in which galaxies trace the underlying dark matter field means that extracting the growth rate from the redshift space clustering can be difficult. In particular, on linear scales the growth rate is exactly degenerate with the galaxy bias, a degeneracy which remains extremely tight even on non-linear scales. Because the peculiar velocity power spectrum relies solely on the growth rate, with no dependence on the galaxy bias (this may not be true in the presence of velocity bias as detailed later in this study). \cite{Koda2014} showed that the combination of redshift space information and peculiar velocities can dramatically improve the growth rate constraints. In this section we present constraints for the datasets in Section~\ref{sec:data} when both the redshifts and peculiar velocities are used. For a given peculiar velocity sample, this information effectively comes `for free' as the redshifts are required for each object anyway. 

The study of \cite{Koda2014} was mainly focused on the case where the number density of redshift tracers was equal to that of the peculiar velocity sample, however due to the signal-to-noise requirements, it is common for the number density of galaxies with measured redshifts to far outweigh that of the peculiar velocities, and to cover much larger cosmological volumes. Hence, in this section we also investigate the constraints on the growth rate when the full redshift sample is used, i.e., we look at the constraints for 6dFGS, TAIPAN and WALLABY+WNSHS where both the full redshift survey and the peculiar velocity subsample are combined, treating the overlapping and non-overlapping regions separately as in Section~\ref{sec:Fisher_2tracer}. We also look at whether the constraints can be improved further by combining the full information from multiple surveys, in particular the addition of the 2MTF data to the 6dFGS redshift and velocity surveys, and the combination of the TAIPAN and WALLABY+WNSHS redshifts and peculiar velocities. The constraints on the growth rate, galaxy bias and nuisance parameters for different combinations of surveys are given in Table~\ref{tab:vel2}, for both $k_{max}=0.1\hompc$ and $k_{max}=0.2\hompc$.

In all cases we find that using the ``free" redshift information in the peculiar velocity samples has a significant effect on the constraints, even though there are more unknown nuisance parameters to take into account. The measurements of the growth rate are improved by $30-50\%$ for all samples, with greater improvement for $k_{max}=0.2\hompc$. This reflects the fact that the velocity field produces stronger constraints on linear scales where the fractional distance error has less effect, whilst including smaller scales has a large effect for the redshift space measurements. We find that our forecasts for the 6dFGSv survey are in good agreement with the forecasts of \cite{Koda2014}, although our constraints for WALLABY+WNSHS and TAIPAN are even tighter as we include the combination of the full redshift survey and the peculiar velocity subsample. The constraints from the 6dFGSv are now slightly better than 2MTF due to its higher number density, which reduces the shot-noise in both the velocity power spectrum and density power spectrum measurements and can no longer be compensated for by the smaller fractional distance error in the 2MTF sample.

\begin{figure}
\centering
\includegraphics[width=0.5\textwidth]{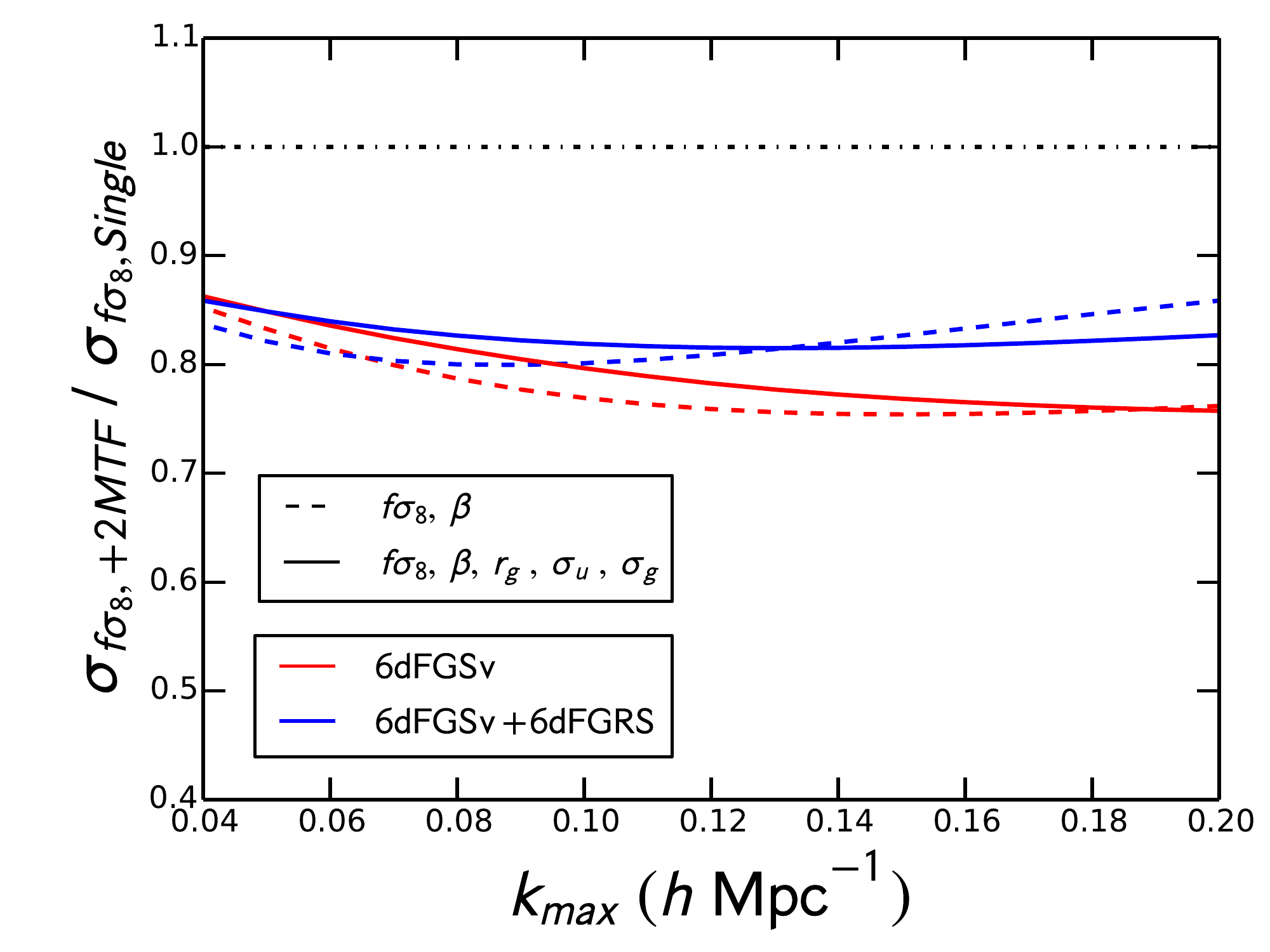}
  \caption{The improvement in the constraints on $f\sigma_{8}$ as a function of $k_{max}$ using information in the velocity and density fields when we add the 2MTF data to the 6dFGSv and 6dFGRS datasets. For the y-axis $\sigma_{f\sigma_{8},Single}$ is the predicted error using either 6dFGSv alone or 6dFGSv+6dFGRS whilst $\sigma_{f\sigma_{8},+2MTF}$ is the error when we add the 2MTF survey i.e., 2MTF+6dFGSv or 2MTF+6dFGSv+6dFGSRS. Hence by plotting the ratio we are directly showing the improvement in the $1\sigma$ errors when we add the 2MTF data to the 6dFGSv data (red) and 6dFGSv+6dFGRS data (blue). Solid and dashed lines show the ratio with and without marginalisation over the nuisance parameters respectively. The black dot-dash line is just to guide the eye, and indicates no improvement from the addition of 2MTF data. For all $k_{max}$ and regardless of our knowledge of the nuisance parameters, we find that the inclusion of the 2MTF data improves the constraints by $\approx 20\%$. This is remarkable when one considers that the number of galaxies being introduced is very small compared to the size of the 6dFGSv+6dFGRS sample, and highlights the constraining power afforded by only a small number of peculiar velocity measurements.}
  \label{fig:fsigma8_improv}
\end{figure}

\begin{figure}
\centering
\includegraphics[width=0.5\textwidth]{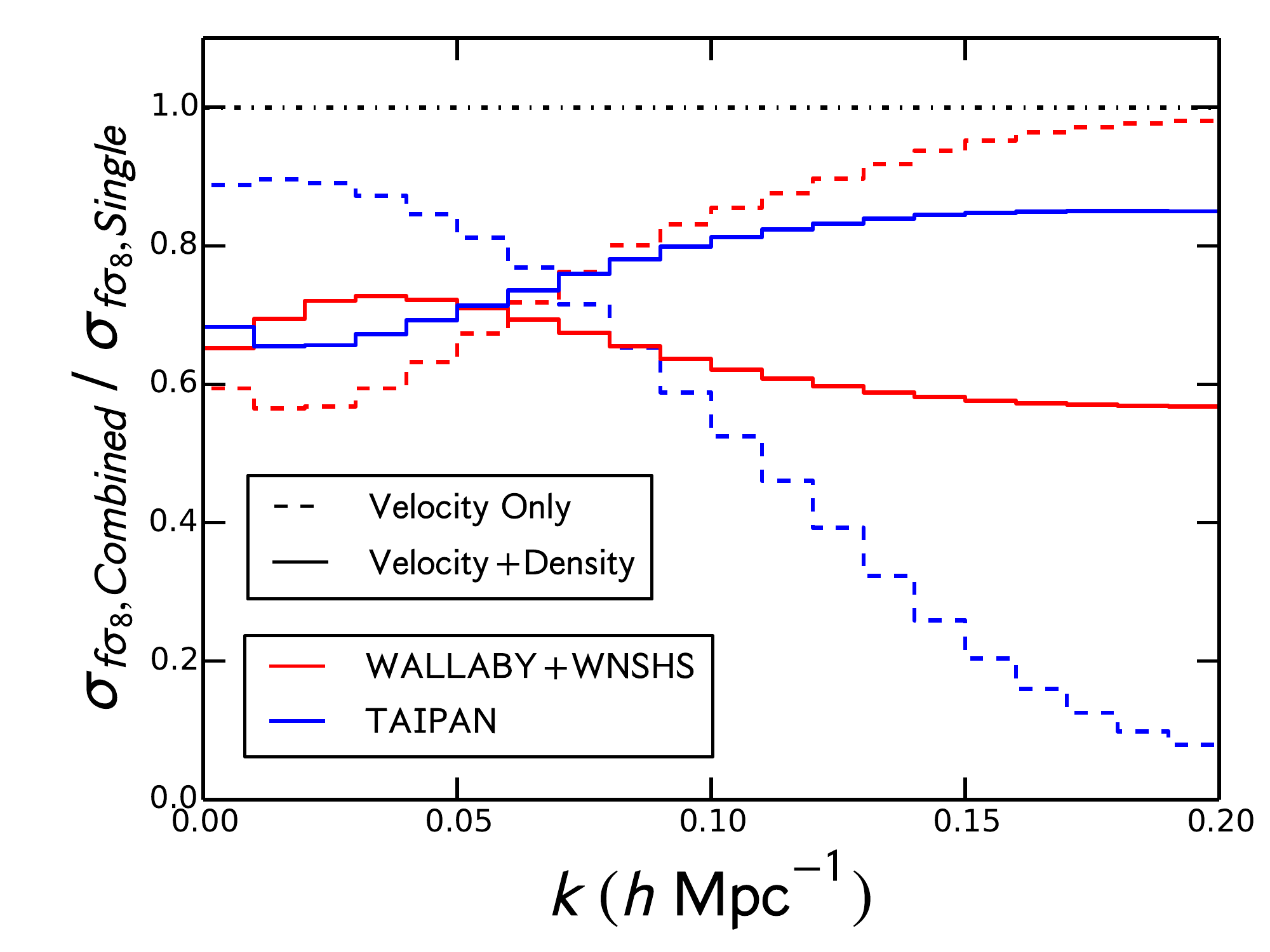}
  \caption{A demonstration of the improvement on \textit{scale-dependent} measurements of $f\sigma_{8}$ when combining the WALLABY+WNSHS and TAIPAN surveys. Similar to Fig.~\ref{fig:fsigma8_improv} $\sigma_{f\sigma_{8},Combined}$ is the $1\sigma$ error for TAIPAN+WALLABY+WNSHS, whilst $\sigma_{f\sigma_{8},Single}$ is for the individual TAIPAN (blue) and WALLABY+WNSHS (red) surveys. Hence, the plot shows the improvement in the errors on $f\sigma_{8}$ when we combine surveys. However, unlike Fig.~\ref{fig:fsigma8_improv}, we are looking at the improvement on the \textit{scale-dependent} measurements of the growth rate measured in bins of width $\Delta k = 0.01\hompc$ and so both $k_{min}$ and $k_{max}$ vary for each bin. Dashed lines show this improvement when using only the velocity field information (the observable $u(\bk)$) whilst the solid lines are using the velocity and density field information. The black dot-dash line is to guide the eye and would indicate no improvement from combining the datasets.}
 \label{fig:fsigma8_sd_improv}
\end{figure}

We find that there are significant gains to be had in combining the full redshift and velocity surveys from 6dFGS, with constraints on the growth rate improving by an additional 30\% compared to the case where we use the redshift information from the peculiar velocity subsample only. We also find that the addition of the 2MTF survey still improves the constraints compared to the 6dFGSv and, surprisingly, the 6dFGSv+6dFGRS sample, even though there are additional nuisance parameters to marginalise over. Figure~\ref{fig:fsigma8_improv} shows the percentage improvement on the 1$\sigma$ errors for $f\sigma_{8}$ as a function of $k_{max}$ when we add the 2MTF data. We find an almost constant 20\% improvement, regardless of the $k_{max}$. This is due to the large overlap area and difference in bias between the 2MTF and 6dFGS, and the constraining power introduced by the additional peculiar velocity measurements. It has already been shown that a small number of peculiar velocity measurements can improve constraints on the growth rate compared to redshift information alone, and adding the 2MTF data gives an additional 20\% peculiar velocities compared to the 6dFGSv sample alone. 

Finally, we see that combining the full WALLABY+WNSHS and TAIPAN surveys also improves the constraints by $\approx 25\%$ compared to the individual surveys. This is similar to the improvement found when combining just the peculiar velocity subsamples, which combined with the result of \cite{Beutler2012} that this combination does little for the constraints using the density field alone, indicates that combining the peculiar velocity samples has a sizeable impact on the statistical power of the samples, without even considering the fact that such a combination would likely improve the systematic robustness of the results too. 

This improvement is also true for scale dependent measurements of the growth rate, as shown in Fig.~\ref{fig:fsigma8_sd_improv}. Here we show the ratio of the errors on the $f\sigma_{8}$ measurement for the combined and separate samples, for both the velocity field only and the combined velocity and density fields, in bins of width $\Delta k=0.01\hompc$. We find an interesting trend for the velocity field, which is that combining the two surveys improves the constraints for the TAIPAN survey mostly on small scales, whilst improving the results from WALLABY+WNSHS alone on large scales. Hence the combination of the two surveys has much greater potential for constraining the scale dependence of the growth rate than either of these surveys individually. The trend is less apparent for the constraints using both the velocity and density fields, although combining these two still improves the individual constraints for every $k$-bin.

Overall we find that the combination of WALLABY+WNSHS and TAIPAN has the ability to achieve a measurement error of between $2\%$ and $3\%$ at $k_{max}=0.2\hompc$ depending on our knowledge of the nuisance parameters. Such a constraint at low redshift will be able to put very tight limits on possible extensions to General Relativity.

\subsection{$f_{NL}$ constraints}

In this section we investigate whether peculiar velocities can be used to improve constraints on primordial, local-type, non-Gaussianity via the parameter $f_{NL}$. We expect the use of the velocity field can partially break the degeneracy with bias and add additional information from the cross spectra between the density and velocity fields. Table~\ref{tab:fnl} shows the predicted $1\sigma$ errors on $f_{NL}$ for the WALLABY+WNSHS and TAIPAN surveys, and their combination, using only the redshift information and with additional information from the peculiar velocity subsample.

For the two surveys alone, we find that the inclusion of peculiar velocities improves constraints by up to $40\%$, with larger gains if we have to marginalise over the bias, as the velocity field helps partially break the degeneracy between $\beta$ and $f_{NL}$. We find that the velocity field gives stronger improvement for the WALLABY+WNSHS survey than for TAIPAN, the latter of which is less sensitive to primordial non-Gaussianity as its galaxy bias is closer to 1. The combination of WALLABY+WNSHS and TAIPAN vastly improves constraints compared to the two surveys alone, by at least a factor of $2$. The constraints on $f_{NL}$ are very sensitive to the use of the multi-tracer technique \mbox{\citep{Mcdonald2009b}} and the large difference between the bias of the TAIPAN and WALLABY+WNSHS galaxies has a large effect. This is especially apparent for these two samples as the additional scale-dependent bias introduced by the primordial non-Gaussianity depends on the factor $b-1$, which acts in opposite directions for the WALLABY+WNSHS and TAIPAN samples. Hence for a given value of $f_{NL}$, there is a large difference between the large scale power spectra. However, we do find that when the two surveys are combined the benefit of using the peculiar velocity subsamples is minimal, as the degeneracy between $\beta$ and $f_{NL}$ is already significantly reduced by the use of two very different tracers of the density field.

Whilst the effect of the multi-tracer technique and peculiar velocity measurements on the $f_{NL}$ constraints is interesting, it is worth noting that for the two samples the constraints on $f_{NL}$ are not competitive compared to other large-scale structure surveys within the same time-frame. For instance, the Fisher matrix forecasts of \cite{Zhao2016} for the extended Baryon Oscillation Spectroscopic Survey (eBOSS, \citealt{Dawson2016}) predict similar constraints using emission line-galaxies, but constraints that are a factor of over $3$ better using Luminous Red Galaxies. Combining all the samples from this survey can give even better constraints ($\sigma_{f_{NL}}\sim 15$). This increased constraining power comes from the much larger cosmological volume probed by the eBOSS survey and the range of different biases probed; due to the $k^{-2}$ dependence of the scale-dependent bias from primordial non-Gaussianity, the constraints on $f_{NL}$ are very sensitive to the largest modes that can be measured.

\begin{table}
\caption{$1\sigma$ errors for $f_{NL}$ for the WALLABY+WNSHS and TAIPAN surveys, separately and in combination, using $k_{max}=0.2\hompc$. We provide predictions for constraints from the density field alone, and when combined with the velocity field assuming perfect knowledge of the galaxy bias and marginalising over it.}
\centering
\begin{tabular}{llccc} \hline
\multicolumn{2}{c}{$f_{NL}$ constraints} & \multicolumn{2}{c}{$\sigma(f_{NL})$}\\
Survey  			& Parameters & Density Only & Density + Velocity \\ \hline
TAIPAN   		& $f_{NL}$ 			& 116.5 & 111.5  \\
	     		& $f_{NL},\,\beta$ 	& 175.6 & 161.1  \vspace{5pt} \\ 
WALLABY+WNSHS   	& $f_{NL}$ 			& 116.7 &  79.1  \\
	     		& $f_{NL},\,\beta$ 	& 183.3 &  97.0  \vspace{5pt} \\ 
TAIPAN+   		& $f_{NL}$ 			&  52.5 &  46.0  \\
WALLABY+WNSHS	& $f_{NL},\,\beta$ 	&  74.5 &  60.3  \vspace{5pt} \\ 

\end{tabular}
\label{tab:fnl}
\end{table}

\subsection{$\gamma$ constraints} \label{sec:gammaconstraints}

In order to investigate the constraints available on the $\gamma$ parameter one could use the expressions in Section~\ref{sec:gamma} to extend the power spectrum models. However, because we are assuming a fixed power spectrum shape a much simpler method is available. The Fisher matrix including the $\gamma$ parameter can be obtained simply by performing a transformation of the Fisher matrix of our fiducial parameter combination. If we have the Fisher matrix $\boldsymbol{F}$ for a set of parameters $\boldsymbol{\lambda} = \{f\sigma_{8}, \beta, r_{g}, \sigma_{g}, \sigma_{u}\}$ then the new Fisher matrix $\boldsymbol{F'}$ for parameters $\boldsymbol{\lambda'} = \{\Omega_{m}, \gamma, \beta, r_{g}, \sigma_{g}, \sigma_{u}\}$ is given by
\begin{equation}
\boldsymbol{F'} = \boldsymbol{M}^{T}\boldsymbol{F}\boldsymbol{M},
\end{equation}
where $\boldsymbol{M}$ is the transformation matrix between the two sets of variables, $M_{ij} = \frac{\partial \lambda_{i}}{\partial \lambda'_{j}}$ \citep{Coe2009}. The only derivatives of interest are those of $f\sigma_{8}$ with respect to $\Omega_{m}$ and $\gamma$, which we evaluate using the previous expressions in Section~\ref{sec:gamma} and our fiducial cosmological parameters with $\gamma=0.55$. We solve the case of $\frac{\partial \sigma_{8}}{\partial \Omega_{m}}$, by finite differencing the values of $\sigma_{8}$ output by {\sc camb} for different $\Omega_{m}$.

\begin{figure}
\centering
\includegraphics[width=0.5\textwidth]{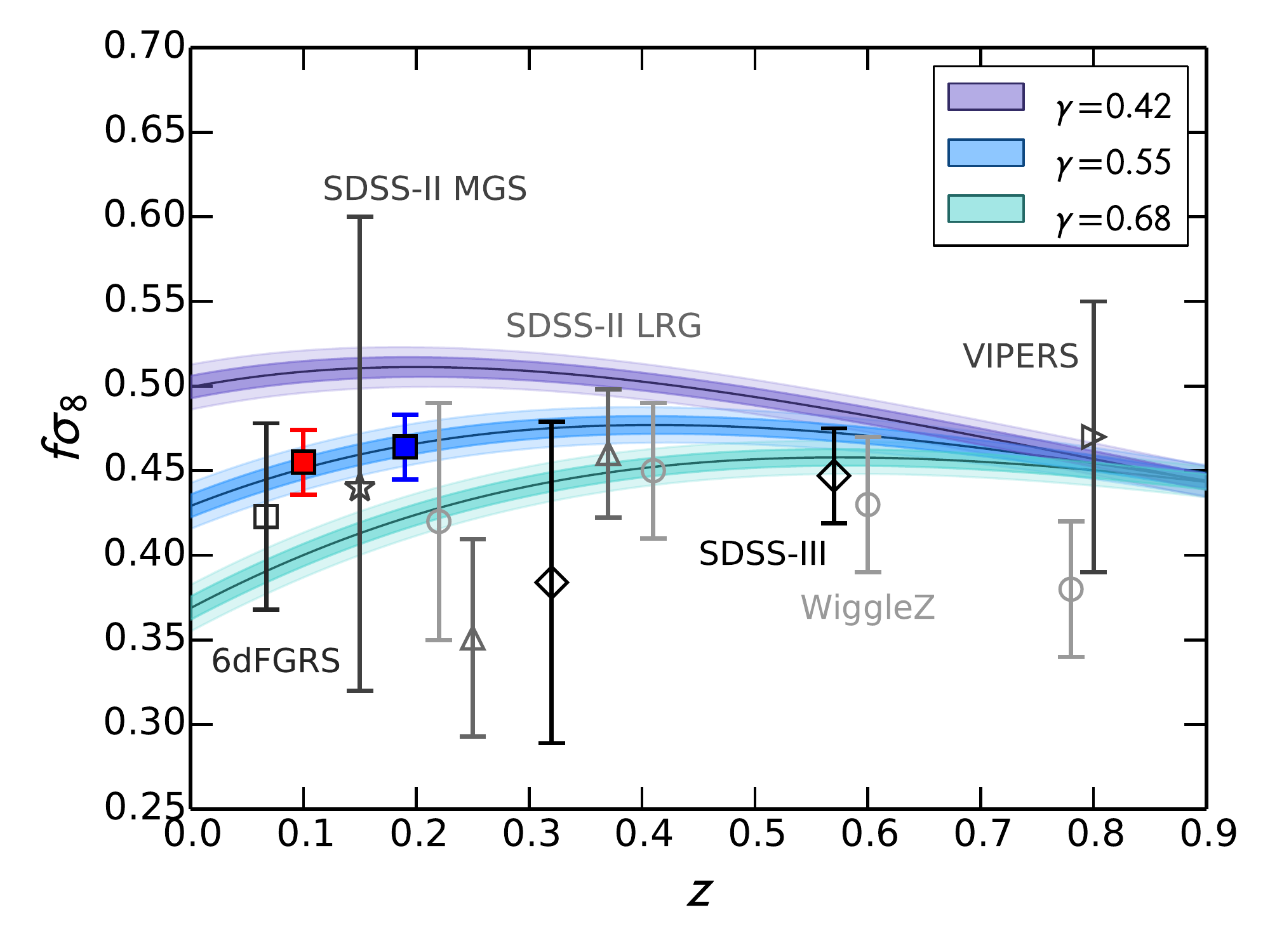}
  \caption{Comparison of measurements of the growth rate from a variety of galaxy surveys. Forecasts for the TAIPAN and WALLABY+WNSHS surveys are shown as filled blue and red squares respectively. Other data points represent the 6dFGRS \citep{Beutler2012}, SDSS-II MGS \citep{Howlett2015}, SDSS-II LRG \citep{Oka2014}, SDSS-III BOSS \citep{Chuang2013,Samushia2014}, WiggleZ \citep{Blake2011a, Blake2011b} and VIPERS \citep{delaTorre2013} surveys. We have also included best-fit lines and $68\%$ and $95\%$ confidence regions on the growth rate for values of $\gamma = 0.42$, $0.55$ and $0.68$ based on the results of \protect\cite{Planck2015a}. These confidence regions are calculated by importance sampling the Planck cosmological chains to incorporate the measurement errors as in \protect\cite{Howlett2015}. For consistency we have preferentially chosen to plot, where possible, results that \textit{do not} include the degeneracy between RSD and the Alcock-Paczynski effect \citep{Alcock1979}, as we have also neglected this in our forecasts. We expect this effect to be small at the low redshift of the TAIPAN and WALLABY+WNSHS samples.}
  \label{fig:fsigma8_comp}
\end{figure}

Without the addition of any extra information, this results in a singular matrix as the two parameters $\Omega_{m}$ and $\gamma$ are completely degenerate. To overcome this, in a procedure that will likely be done for future measurements anyway, we add a Gaussian prior on $\Omega_{m}$ of width $\sigma_{\Omega_{m}}=0.0062$ based on CMB measurements \citep{Planck2015a}.

Constraints on $\gamma$ for all the surveys considered in this paper are shown in Table~\ref{tab:gamma}. We show the predicted percentage error for the velocity field only, and the combination of velocity and density field information, for $k_{max}=0.2\hompc$, marginalising over all nuisance parameters. Again we find that the addition of the 2MTF data improves the constraints on $\gamma$ by $15-30\%$, even over the case where we combined all the data from the 6dFGS. The constraints from the velocity field alone for 2MTF and 6dFGSv are comparable to the constraints from the MGS redshift sample of $\sim 63,000$ galaxies used by \cite{Howlett2015}, highlighting the strong constraining power offered by the peculiar velocity measurements. These are significantly improved by the inclusion of the ``free" density field information from these datasets. The results from the 6dFGSv+6dFGRS data are in good agreement with the results of \cite{Beutler2012} obtained by analysing the redshift space clustering of the 6dFGRS (a $16\%$ measurement of $\gamma$), although the inclusion of the velocity subsample does improve the constraints slightly.

\begin{table}
\caption{Fisher matrix forecasts for the percentage uncertainties on $\gamma$ for current and next generation peculiar velocity surveys for $k_{max}=0.2\hompc$, marginalising over all other nuisance parameters, and using information in the velocity field only, or in both the velocity and density fields.}
\centering
\begin{tabular}{llccc} \hline
$\gamma$ constraints & \multicolumn{2}{c}{$100\times\sigma(\gamma)\,/\,\gamma$}\\
Survey					& Velocity Only & Velocity + Density \\ \hline
2MTF    					& 40.4 & 24.0 \vspace{3pt} \\
6dFGSv    				& 37.4 & 20.3 \vspace{3pt}\\
6dFGSv + 6dFGRS    		& 37.4 & 13.6 \vspace{3pt}\\
2MTF + 6dFGSv    		& 28.4 & 15.5 \vspace{3pt} \\
2MTF + 6dFGSv + 6dFGRS   	& 28.4 & 11.3 \vspace{3pt} \\
TAIPAN    				& 15.2 &  5.2 \vspace{3pt} \\
WALLABY + WNSHS    		& 16.4 &  5.3 \vspace{3pt} \\
TAIPAN + WALLABY + WNSHS 	& 11.5 &  4.0 \vspace{3pt} \\

\end{tabular}
\label{tab:gamma}
\end{table}

Using the velocity and density fields for the TAIPAN and WALLABY+WNSHS surveys predicts very tight constraints on the value of $\gamma$, which will provide a very strong consistency test of GR. For comparison \cite{Samushia2014} found a 16\% measurement of $\gamma$ using BOSS-DR11 and CMB data, which can be matched by the peculiar velocity measurements from these surveys alone. The low redshift of the WALLABY+WNSHS and TAIPAN samples allow for a much more stringent test of deviations from General Relativity, as it is in this regime where different values of $\gamma$ can produce the widest divergence in the growth rate of structure. This is highlighted in Fig.~\ref{fig:fsigma8_comp} where we plot a range of $f\sigma_{8}$ measurements at different redshifts from different studies against the predictions for different values of $\gamma$ using a prior on $\Omega_{m}$ from the CMB \citep{Planck2015a}. Also plotted are the predicted $f\sigma_{8}$ constraints for the WALLABY+WNSHS and TAIPAN surveys at $z\approx0.1$ and $z\approx0.2$ respectively.

\section{Systematic tests} \label{sec:systematics}

In this section we explore the effects that potential systematics may have on growth rate measurements obtained with the next generation TAIPAN and WALLABY+WNSHS surveys. In particular we look at the potential effects of scale dependent spatial and velocity bias and offsets in the zero-point. It should be noted that whilst reasonable values have been adopted for tests in this Section, the strength of any systematic effects will depend strongly on these values. The purpose of this Section is merely to highlight possible systematics that should be taken into consideration when modelling next generation redshift and peculiar velocity surveys, but for a given survey the magnitude of these effects may differ from those presented here.

\begin{table*}
\caption{A table detailing the effects of scale-dependent galaxy bias on the $f\sigma_{8}$ constraints for the TAIPAN and WALLABY+WNSHS surveys. The second and third columns give the systematic offset of $f\sigma_{8}$ and $\beta$ respectively from the fiducial value as a percentage of the $1\sigma$ errors, i.e., a value of $100\%$ indicates that neglecting scale-dependent galaxy bias shifts the constraints on that parameter by $1\sigma$ away from the true value. The fourth and fifth columns gives the percentage error on $f\sigma_{8}$ with and without marginalisation over the value of $b_{\zeta}$. The final column gives the percentage constraints on $b_{\zeta}$ itself. All results assume perfect knowledge of other nuisance parameters, only $f\sigma_{8}$ and $\beta$ are left free, and we set $k_{max}=0.2\hompc$. The top half of the table gives the results when only the redshifts are used to measure the density field, whilst the lower half shows results from combining the redshifts with the peculiar velocities to measure the velocity and density fields.}
\centering
\begin{tabular}{lccccc} \hline
Survey  & \multicolumn{2}{c}{Percentage Bias ($f\sigma_{8}$, $\beta$)}  & $100\times\sigma(f\sigma_{8})\,/\,f\sigma_{8}$ & $100\times\sigma(f\sigma_{8})\,/\,f\sigma_{8}$ Marginalised & $100\times\sigma(b_{\zeta})\,/\,b_{\zeta}$ \\ \hline
\multicolumn{6}{r}{Density Field Only} \\
TAIPAN   			&  31.8 & -421.9 &  2.6 &  2.6 &  4.5  \vspace{3pt} \\
WALLABY+WNSHS		&   2.4 &  -87.9 &  3.8 &  3.8 & 35.9  \vspace{3pt} \\
\multicolumn{6}{r}{Velocity + Density Fields}   \\
TAIPAN   			&  91.3 & -402.4 &  2.3 &  2.4 &  4.5  \vspace{3pt} \\
WALLABY+WNSHS		&  48.8 &  -75.6 &  2.7 &  2.8 & 35.2  \vspace{3pt} \\
\end{tabular}
\label{tab:sd_bias}
\end{table*}

\subsection{Fisher matrix bias formalism}

\cite{Huterer2005}, \cite{Huterer2006} and \cite{Amara2007} present a simple expression under the Fisher matrix formalism that allows for quantification of the offsets of parameters from their fiducial values due to systematic bias. This offset, $\Delta \theta_{i}$, can be given by projecting the inverse of the true Fisher matrix $\boldsymbol{F}$ along some bias vector $\boldsymbol{B}$
\begin{equation}
\Delta \theta_{j} = \sum_{i} (F^{-1})_{ij}B_{i}
\label{eq:paramshift}
\end{equation}
where the bias vector is computed in a similar way to the Fisher matrix,
\begin{equation}
B_{i} = \frac{1}{2}\int \frac{d^{3}xd^{3}k}{(2\pi)^{3}}\,\mathrm{Tr}\left[\bC^{-1}\frac{\partial{\bC}}{\partial{\lambda_{i}}}\bC^{-1}(\tilde{\bC}-\bC)\right].
\end{equation}
$\bC$ is the true covariance matrix and $\tilde{\bC}$ is the systematically biased covariance matrix. When we look at the effect of neglecting scale-dependent bias, we compute the true covariance matrix and Fisher matrix using model power spectra that include the scale-dependence, whilst the biased covariance matrix is computed using only linear galaxy bias, or a velocity bias of unity. When looking at the effects of a zero-point offset, the true covariance is that without any offset.

The parameter bias can then be compared to the expected error on the parameter to gauge its significance, i.e., by looking at the systematic shift in the parameter given by Eq.~\ref{eq:paramshift} as a percentage of the $1\sigma$ errors given by the usual Fisher matrix calculation. However, the presence of the systematic error will also have an effect on the Fisher matrix forecasts and subsequent parameter errors. Hence when comparing the systematic bias to the error we use the Fisher matrix forecasts including the systematic effect, i.e., using the model without velocity bias.

\subsection{Scale-dependent galaxy bias}

\begin{figure}
\centering
\includegraphics[width=0.5\textwidth]{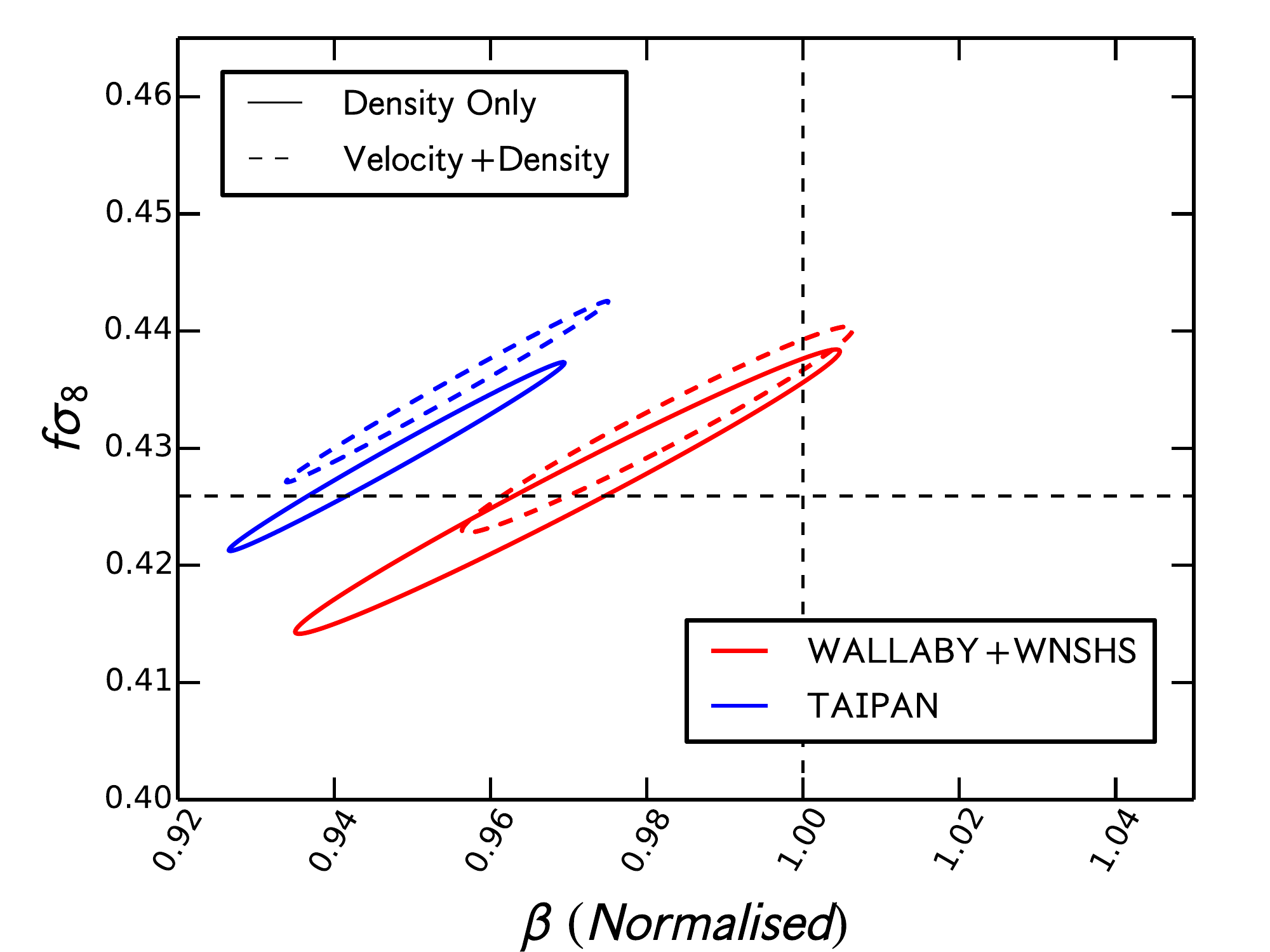}
  \caption{Predicted 68\% confidence ellipses on $f\sigma_{8}$ and $\beta$ (normalised to 1 for both surveys by dividing out the fiducial values from Section~\ref{sec:data}) neglecting scale-dependent spatial bias for the TAIPAN and WALLABY+WNSHS surveys. The predictions using only the density field and using both the velocity and density field are shown as solid and dashed contours respectively. The black dashed lines show our fiducial values for the growth rate and galaxy bias.}
   \label{fig:sd_bias}
\end{figure}

The above formalism is first used to investigate the effect of neglecting scale-dependent galaxy bias on measurements of the growth rate. The predicted systematic bias on the growth rate as a percentage of the $1\sigma$ errors for the TAIPAN and WALLABY+WNSHS surveys is given in Table~\ref{tab:sd_bias}. We find that results using just the density field are subject to only a small ($<0.2\sigma$) shift in the values of $f\sigma_{8}$ from their fiducial values. As also shown in Table~\ref{tab:sd_bias} any systematic bias is largely absorbed into the value of the $\beta$ and the linear galaxy bias, which is generally treated as a nuisance parameter in measurements of the growth rate. This is shown graphically as $1\sigma$ confidence ellipses in Fig~\ref{fig:sd_bias}, where we have normalised $\beta$ to 1 for both surveys by dividing out the fiducial values from Section~\ref{sec:data}.

When information from the velocity field is also included, the potential for systematically biased constraints on the growth rate increases, reaching $~1\sigma$ for the TAIPAN survey. Combining the velocity and density fields breaks the degeneracy between the growth rate and galaxy bias, such that the systematic effects of neglecting the scale-dependence of the galaxy bias are not able to be compensated for as easily as for as when only the density field is used.

Also included in Table~\ref{tab:sd_bias} are the constraints on $f\sigma_{8}$ with and without marginalising over the scale-dependent galaxy bias parameter $b_{\zeta}$, and the constraints on this parameter itself. We find that even though neglecting it can cause systematic shifts, marginalising over $b_{\zeta}$ actually has little effect on the constraints on the growth rate as a majority of the information on the growth rate comes from linear scales. Similar results were found in \cite{Desjacques2010a}. 

\begin{table*}
\caption{A table detailing the effects of velocity bias on the $f\sigma_{8}$ constraints for the TAIPAN and WALLABY+WNSHS surveys. The third column gives the systematic offset of $f\sigma_{8}$ from the fiducial value as a percentage of the 1$\sigma$ errors. The fourth and fifth columns gives the percentage error on $f\sigma_{8}$ with and without marginalisation over the value of $R_{v}$, for $k_{max}=0.2\hompc$. The final column gives the percentage error on $R_{v}$ itself. The top half of the table gives the results when only the peculiar velocities and the velocity field are used, whilst the lower half shows results from combining the velocity and density field information. For each survey we consider the case with and without marginalisation over the other nuisance parameters $r_{g}$, $\sigma_{g}$ and $\sigma_{u}$, with the parameters we include in the Fisher matrix calculation given in the ``Parameters'' column}
\centering
\begin{tabular}{llcccc} \hline
Survey  			& Parameters & Percentage Bias & $100\times\sigma(f\sigma_{8})\,/\,f\sigma_{8}$ & $100\times\sigma(f\sigma_{8})\,/\,f\sigma_{8}$ Marginalised & $100\times\sigma(R_{v})\,/\,R_{v}$ \\ \hline
\multicolumn{6}{r}{Velocity Field Only} \\
TAIPAN   		& $f\sigma_{8}$ 				& -26.3 &  9.9 & 14.0 &  192.7 \\
	     		& $f\sigma_{8},\,\sigma_{u}$ 	&  -0.6 & 13.7 & 16.1 & 2821.6 \vspace{5pt} \\ 
WALLABY+WNSHS   	& $f\sigma_{8}$ 				& -13.9 & 10.8 & 15.9 &  400.0 \\
	     		& $f\sigma_{8},\,\sigma_{u}$ 	&  -2.0 & 14.2 & 18.6 & 2027.7 \vspace{5pt} \\ 
\multicolumn{6}{r}{Velocity + Density Fields} \\
TAIPAN   		& $f\sigma_{8},\,\beta$ 									& -589.8 &  2.6 &  3.6 &   9.4 \\
	     		& $f\sigma_{8},\,\beta,\,r_{g},\,\sigma_{u},\,\sigma_{g}$  &   33.3 &  4.1 &  4.9 & 111.3 \vspace{5pt} \\ 
WALLABY+WNSHS   	& $f\sigma_{8},\,\beta$ 									&  -92.7 &  2.8 &  4.0 &  57.2 \\
	     		& $f\sigma_{8},\,\beta,\,r_{g},\,\sigma_{u},\,\sigma_{g}$  &    0.7 &  4.2 &  4.2 & 369.8 \vspace{5pt} \\ 

\end{tabular}
\label{tab:vel_bias}
\end{table*}

Rather than being just a nuisance parameter, the amplitude of the scale-dependent bias may also contain interesting insight into the link between galaxies and the underlying dark matter. In this sense it is interesting to note that WALLABY+WNSHS, and the TAIPAN survey especially, can be expected to produce strong constraints on this scale dependence, with our forecasts predicting $\sim36\%$ and $\sim5\%$ measurements of $b_{\zeta}$ for these two datasets. Whilst the use of velocity field information improves the constraints on the growth rate and linear galaxy bias, it does little to improve the errors on $b_{\zeta}$ as the velocity power spectrum is primarily noise-dominated on quasi- and non-linear scales.

Whilst we have used a physically motivated model for the form and normalisation of the scale-dependent bias, the exact strength of the systematic effects when neglecting the scale-dependence will be model-dependent. Nonetheless, these predictions motivate more detailed study of the potential systematics due to scale-dependent galaxy bias that could arise in these surveys. This could be done using simulations that mimic the proposed galaxy distributions. Overall, we suggest that a model accounting for scale dependence in the galaxy bias should be used for the next generation TAIPAN and WALLABY+WNSHS surveys, especially as doing so seems to come at little cost to the growth rate constraints.

\subsection{Velocity bias}

We also investigate the systematic effects of neglecting velocity bias on the TAIPAN and WALLABY+WNSHS surveys by calculating the systematic bias on $f\sigma_{8}$ as a percentage of the 1$\sigma$ forecasted errors. The results are summarised in Table~\ref{tab:vel_bias} considering the case where we use just the velocity field, and when the velocity and density fields are combined, with $k_{max}=0.1\hompc$ and $0.2\hompc$.

When using only the information in the velocity field and not including RSD damping effects in our models, we find that we can expect the constraints on $f\sigma_{8}$ to be biased towards lower values by up to $30\%$ for the TAIPAN and WALLABY+WNSHS surveys at $k_{max}=0.2\hompc$, with some reduction in the systematic bias for lower $k_{max}$. The $k^{2}$ dependence of the velocity bias means that the reduction in power is more prevalent on small scales and the effect of neglecting the velocity bias becomes increasingly important as we include smaller scale information.

\begin{figure}
\centering
\includegraphics[width=0.5\textwidth]{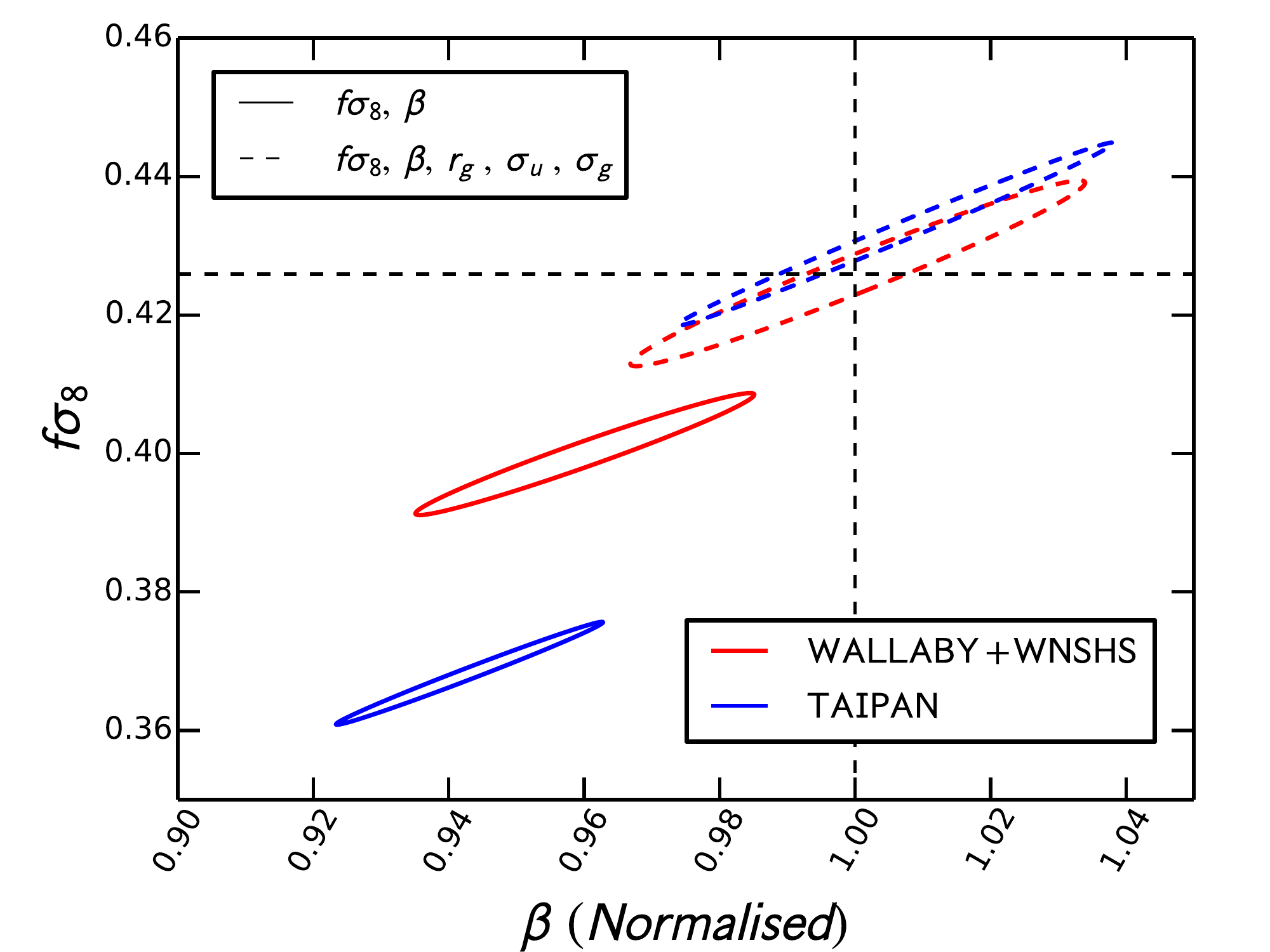}
  \caption{68\% forecasted confidence ellipses on $f\sigma_{8}$ and $\beta$ (normalised to 1) for the TAIPAN and WALLABY+WNSHS surveys when velocity bias is neglected. Our fiducial values are denoted by the dashed black lines. Neglecting velocity bias significantly shifts the measured values away from their fiducial ones when we do not marginalise over the nuisance parameters $r_{g}$, $\sigma_{g}$ and $\sigma_{u}$ (solid contours). This can be compensated for if these nuisance parameters are marginalised over (dashed contours).}
   \label{fig:vel_bias2}
\end{figure}

\begin{figure}
\centering
\includegraphics[width=0.5\textwidth]{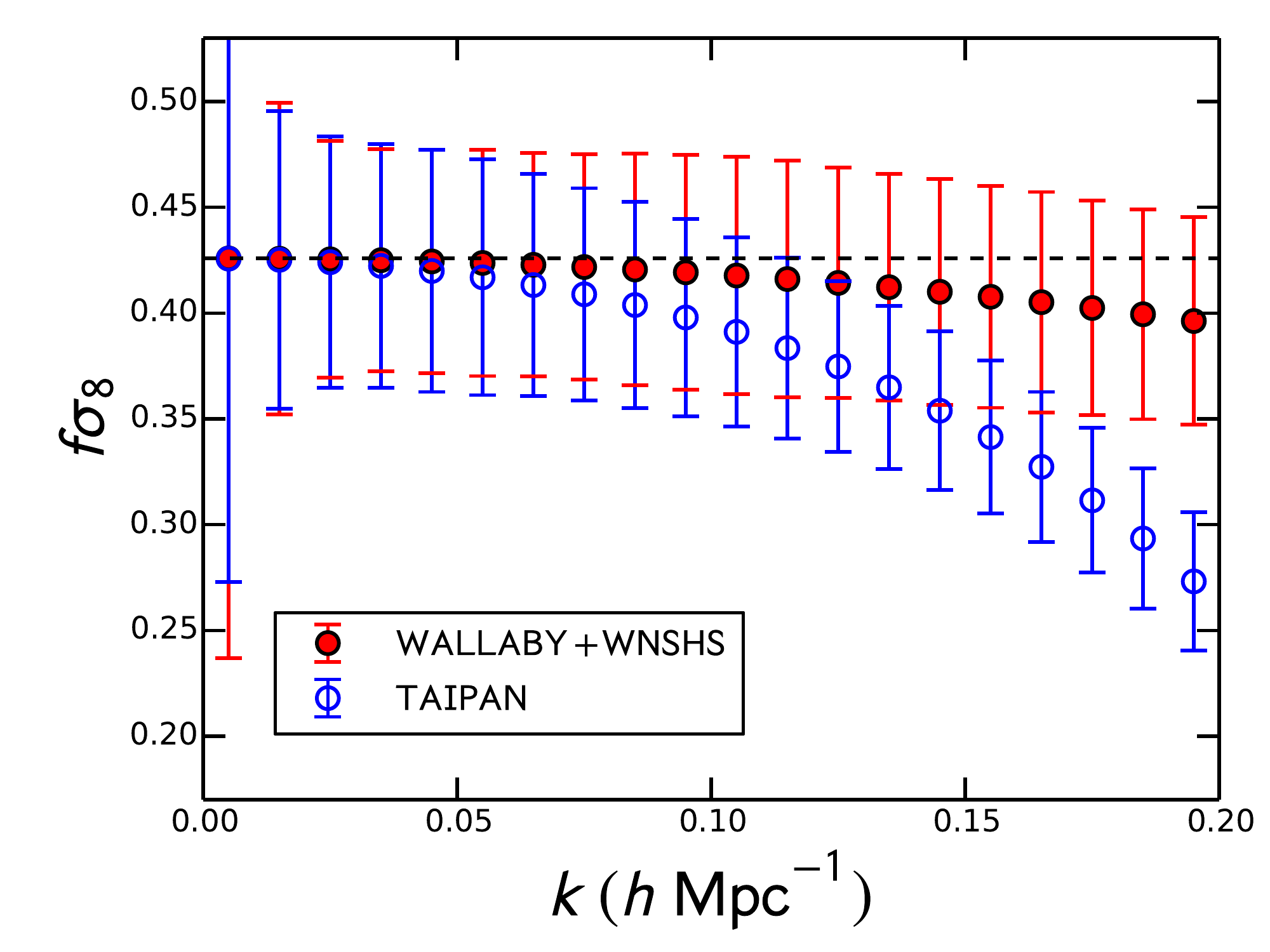}
  \caption{$1\sigma$ constraints on $f\sigma_{8}$ in $k$-bins of width $\Delta k=0.01\hompc$ for the TAIPAN and WALLABY+WNSHS surveys when neglecting the effect of velocity bias. Even though the true growth rate assumed here is scale-independent (dashed line), the forecasts for TAIPAN in different $k$-bins show a strong scale-dependence due to the $k^{2}$ dependence of the velocity bias.}
   \label{fig:vel_bias_sd}
\end{figure}

Interestingly, we find that including and marginalising over the RSD damping can effectively remove this bias on the $f\sigma_{8}$ constraints as the damping of the power spectrum on small scales due to velocity bias is similar to that caused by non-linear RSD. The presence of velocity bias can be compensated for by assuming a stronger non-linear damping term (larger $\sigma_{u}$). This means that the value of $\sigma_{u}$ is significantly biased whilst the constraints on $f\sigma_{8}$ become unbiased. In the sense that $\sigma_{u}$ is treated as a nuisance parameter, this means that we can neglect the effects of velocity bias so long as a model with appropriate freedom in the non-linear RSD damping is used. This assumption breaks down however if a strong prior is placed on the value of $\sigma_{u}$ (in which case, in the presence of velocity bias, the data will likely prefer a value of $\sigma_{u}$ outside the prior range), or if $\sigma_{u}$ is not treated solely as a nuisance parameter, and one wishes to obtain information on, for instance, the non linear relationship between galaxies and their host dark matter halos. In this case it becomes important to marginalise over the velocity bias.

\begin{table*}
\caption{A table detailing the effects of a zero-point offset on the $f\sigma_{8}$ constraints for the TAIPAN and WALLABY+WNSHS surveys. The third column gives the systematic offset from the fiducial value as a percentage of the $1\sigma$ error whilst the fourth and fifth give the percentage error on $f\sigma_{8}$ with and without marginalisation over the value of $\sigma_{m}$. The final column is the percentage error on the constraints of $\sigma_{m}$ itself. All results are for $k_{max}=0.2\hompc$ and assuming perfect knowledge of nuisance parameters $r_{g}$, $\sigma_{g}$ and $\sigma_{u}$. The top half of the table gives the results when only the peculiar velocity measurements are used, whilst the lower half shows results from combining the velocity and density fields information.}
\centering
\begin{tabular}{llcccc} \hline
Survey  & Parameters & Percentage Bias  & $100\times\sigma(f\sigma_{8})\,/\,f\sigma_{8}$ & $100\times\sigma(f\sigma_{8})\,/\,f\sigma_{8}$ Marginalised & $100\times\sigma(\sigma_{x})\,/\,\sigma_{x}$ \\ \hline
\multicolumn{6}{r}{Velocity Field Only} \\
TAIPAN   		& $f\sigma_{8}$ 				& 10.6 &  9.9 & 10.0 & 79.7 \vspace{3pt} \\
WALLABY+WNSHS   	& $f\sigma_{8}$ 				&  6.3 & 10.7 & 10.7 & 23.7 \vspace{3pt} \\
\multicolumn{6}{r}{Velocity + Density Fields} \\
TAIPAN   		& $f\sigma_{8},\,\beta$ 		&  0.4 &  2.3 &  2.3 & 77.3 \vspace{3pt}\\
WALLABY+WNSHS   	& $f\sigma_{8},\,\beta$ 		&  0.1 &  2.7 &  2.7 & 23.6 \vspace{3pt} \\

\end{tabular}
\label{tab:zp_bias}
\end{table*}

Also shown in Table~\ref{tab:vel_bias} is the effect of this marginalisation on the constraints on $f\sigma_{8}$. We find that marginalising over the $R_{v}$ parameter has a large effect on the recovered $1\sigma$ errors for $f\sigma_{8}$, increasing them by between $20\%$ and $40\%$. Again this is in agreement with the results of \cite{Desjacques2010a}.

When combining the density and velocity fields, the systematic effects of velocity bias are much more apparent. The presence of velocity bias affects not just the velocity-velocity power spectrum but also the density-density power spectrum and the cross spectrum between the two fields, such that the systematic effects are increased whilst the statistical errors on $f\sigma_{8}$ are reduced by the large increase in information. We find that when fitting the model assuming perfect knowledge of the nuisance parameters the presence of velocity bias can bias the value of $f\sigma_{8}$ away from its fiducial value by nearly 1$\sigma$ for the WALLABY+WNSHS survey and over $5\sigma$ for the TAIPAN survey at $k_{max}=0.2\hompc$. This is further demonstrated in Fig.~\ref{fig:vel_bias2}, where we plot the 68\% confidence regions on $f\sigma_{8}$ and $\beta$ (normalised to 1 for both surveys) with and without including the effects of velocity bias in the model. We can see that neglecting velocity bias reduces the measured value of $f\sigma_{8}$ far from its true value.

The scale-dependent nature of velocity bias means that not marginalising over this effect can result in measurements of the growth rate that can appear to be scale-dependent and hence seem to be a signature of modified gravity models. In particular if the growth rate is measured in $k$-bins as per Fig.~\ref{fig:fsigma8_sd_improv}, then a scale-\textit{independent} growth rate can still appear scale-dependent in the presence of velocity bias, to high significance for future surveys. 
In Fig.~\ref{fig:vel_bias_sd} we plot the forecasted constraints on $f\sigma_{8}$ for the TAIPAN and WALLABY+WNSHS surveys in $k$-bins of width $\Delta k=0.01\hompc$, which is a similar procedure to that used in \cite{Macaulay2012} and \cite{Johnson2014} to look for scale-dependence. We see that ignoring the $k$ dependence of the velocity bias causes a significant scale-dependence to be measured in the growth rate which could be easily taken as a sign of modified gravity.

\subsection{Zero-point offsets}

The final systematic we test is the effect of an offset between the measured and true value of the zero-point of the astrophysical relation used to infer each galaxy's true distance. A zero-point offset acts as a shot-noise term, raising/lowering the overall amplitude of the velocity power spectrum. The derivation of this effect on the velocity power spectrum is given in Section~\ref{sec:zp}. The effect of a zero-point offset on our predicted measurements of the growth rate is presented in Table~\ref{tab:zp_bias}.

Even though we adopt a significantly higher value for the zero-point offset than would be expected for the next generation surveys we consider (again see Section~\ref{sec:zp}), the effect on the growth rate constraints is very small, being at worst only $10\%$ of the $1\sigma$ errors. The reason is as follows. The zero-point offset acts as a shot-noise term in the velocity power spectrum. However this is also true for the observational errors in the peculiar velocities arising from intrinsic scatter in the astrophysical relation and random non-linear motions.

\begin{figure}
\centering
\includegraphics[width=0.5\textwidth]{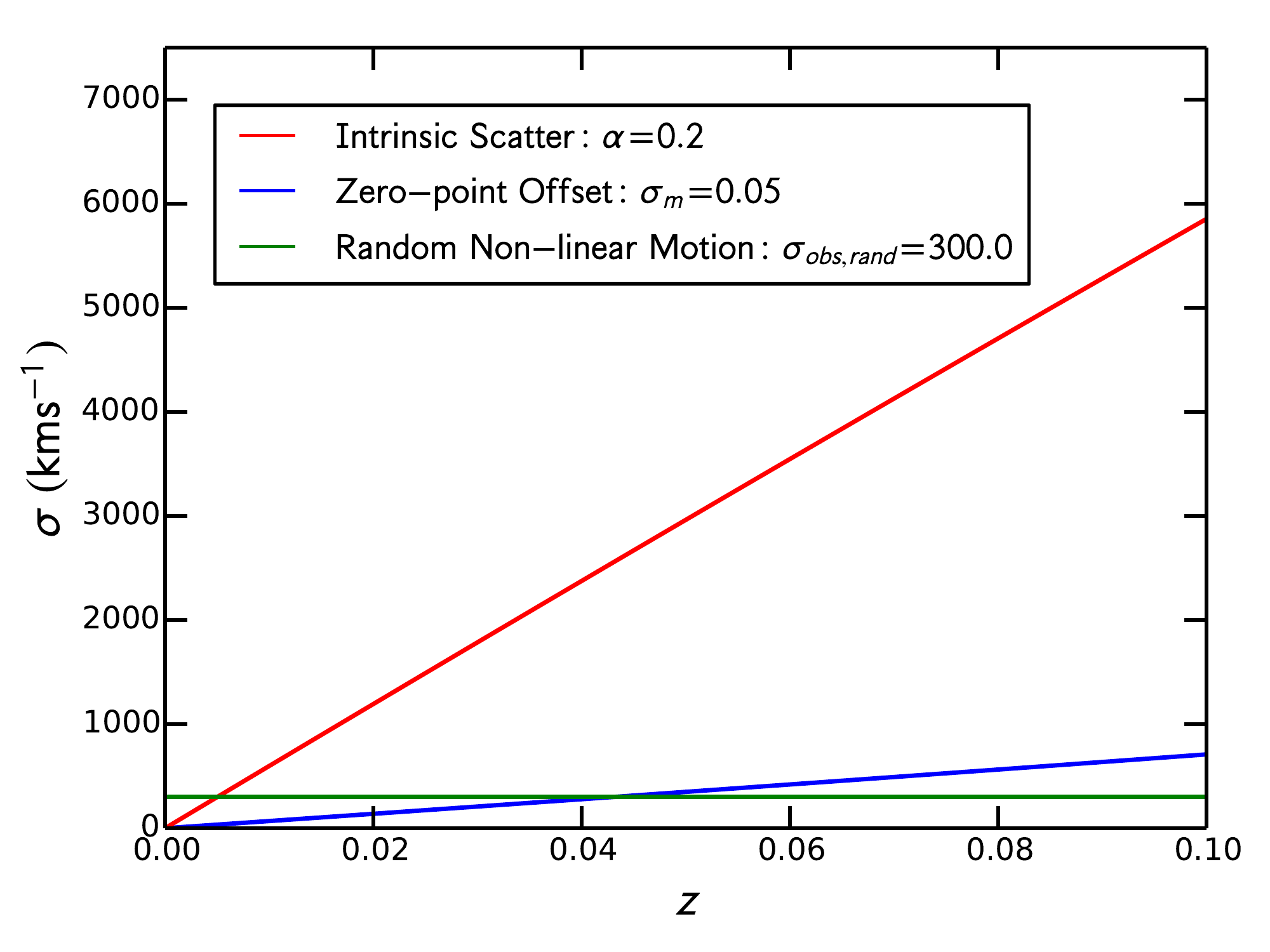}
  \caption{The standard deviation in the line-of-sight peculiar velocity as a function of redshift arising due to the intrinsic scatter in the astrophysical relation (red), random non-linear motions of galaxies (green) and a systematic error in the zero-point offset (blue). These are calculated as in Eqs.~\ref{eq:err} and~\ref{eq:zperr} respectively. For this figure we use reasonable values of $\alpha=0.2$ for the percentage error due to intrinsic scatter, a constant random non-linear velocity dispersion of $\sigma_{obs,rand}=300\,\mathrm{kms^{-1}}$ and an offset in the zero-point reference magnitude of $\sigma_{m}=0.05\,\mathrm{dex}$. At all redshifts the statistical errors due to intrinsic scatter and non-linear velocity dispersion dominate over the systematic error due to a zero-point. As these affect measurements of the velocity power spectrum in the same way, a zero-point offset is hence unlikely to bias cosmological constraints.}
  \label{fig:mag_vel_conv}
\end{figure}

The component due to random non-linear motions is assumed not to vary with redshift, and so can be expected to become subdominant compared to the other contributions above some redshift. The balance then is between the statistical observational error and the systematic error introduced by the zero-point, both of which will increase with redshift. However, looking at the expressions for these two components (Eqs.~\ref{eq:err} and~\ref{eq:zperr}), it becomes apparent that unless the change in the reference magnitude at which the peculiar velocity is zero due to systematic error is significantly larger than that assumed here (which is already very conservative); or the intrinsic scatter in the astrophysical relationship used to infer the galaxy's true distance is much smaller than is currently believed feasible, then the statistical observational error will always dominate over the systematic error from a zero-point offset. In other words, for all reasonable choices of offset in the apparent magnitude $\sigma_{m}$ and fractional error in the distance to a galaxy $\alpha$
\begin{equation}
\alpha H_{0}r(z)\,\gg\,\frac{c\,\mathrm{ln} 10}{5}\left(1 - \frac{c\,(1+z)}{H(z)r(z)}\right)^{-1}\sigma_{m}.
\end{equation}
This is more obvious at very low redshift where we can use the approximation for the comoving distance $r(z) = cz/H_{0}$ and $H(z)=H_{0}$, in which case we arrive at
\begin{equation}
\alpha\,\gg\,\frac{\mathrm{ln} 10}{5}\sigma_{m}.
\end{equation}
This is further shown in Fig.~\ref{fig:mag_vel_conv}, where we plot the different error contributions as a function of redshift for the parameter values assumed in this study, $\alpha=0.2$, $\sigma_{m}=0.05\,\mathrm{dex}$ and $\sigma_{obs,rand}=300\,\mathrm{kms^{-1}}$.

With this in mind we conclude that for measurements of the velocity power spectrum it is highly unlikely that constraints on the growth rate will be biased if an offset in the zero-point is neglected. This is corroborated by the study of \cite{Johnson2014} who found that their results were largely insensitive to whether or not the zero-point was marginalised over. 

That said, looking back at Table~\ref{tab:zp_bias}, marginalising over a possible zero-point offset has negligible impact on the growth rate constraints, whilst it may allow for broad constraints to be placed on the zero-point offset itself. Though we have shown that a zero-point offset hardly affects power spectrum constraints, the same cannot be said of other techniques using peculiar velocity measurements to obtain cosmological information, such as the bulk flow. If the zero-point offset \textit{is} marginalised over, this could be used as a prior for other measurements.

\section{Conclusions} \label{sec:conclusion}

In this work we have used the Fisher matrix formalism to investigate the cosmological constraints that can be obtained using current and next generation peculiar velocity and redshift surveys. This has built on the work of \cite{Burkey2004} and \cite{Koda2014} though with substantial extensions. Our main conclusions can be summarised as follows:
\begin{itemize}
\item{We have extended the redshift space power spectrum models of \cite{Koda2014} to incorporate the effects of primordial non-Gaussianity, scale-dependent biases, and a zero-point offset, as well including the $\gamma$ parameterisation to test the consistency of GR.}\
\item{We have demonstrated how the Fisher matrix approach can be used to forecasts constraints for two surveys each with measurements of the velocity and density field, treating the overlapping and non-overlapping regions of the survey, both in angular and radial directions, as sub-matrices. This method could easily be extended to look at a larger number of peculiar velocity surveys or alternative sets of observables.}
\item{Forecasts on the growth rate have been obtained for the currently available 2MTF and 6dFGS surveys. In particular we find that:
\begin{enumerate}
\item{The peculiar velocity measurements from 2MTF should be able to obtain an independent measurement of the growth rate of structure at $z\approx0$ with accuracy comparable to the 6dFGSv subsample.} \vspace{3pt}
\item{The addition of peculiar velocity and density field measurements from the same survey significantly improves the growth rate constraints, corroborating the results of \cite{Burkey2004} and \cite{Koda2014}. In particular the use of density field measurements from the full 6dFGRS sample, as opposed to only those from the 6dFGSv subsample, also tightens the growth rate constraints noticeably.} \vspace{3pt}
\item{The addition of the 2MTF data to either the 6dFGSv subsample alone, or the full 6dFGS dataset, improves the growth rate constraints by $\sim20\%$ regardless of whether we consider measurements of only the velocity field or combined information from the velocity and density field. This highlights the potential of only a few peculiar velocity measurements to provide strong constraints on the growth rate.}
\end{enumerate}}
\item{Our predictions for both the next generation TAIPAN and WALLABY+WNSHS surveys show that they have the potential to provide extremely tight constraints on the growth rate of structure, which will allow for some of the most stringent tests of modified gravity from growth rate measurements to date. The combination of these two surveys has the potential to measure $f\sigma_{8}$ to $<3\%$.}
\item{Although the use of velocity field information can improve the constraints on primordial non-Gaussianity from the power spectrum, neither the TAIPAN nor WALLABY+WNSHS surveys will be able to provide competitive constraints compared to other next generation surveys on the same time-scale as the cosmological volume probed by these surveys is not large enough.}
\item{Using the Fisher matrix method, we have shown that both scale-dependent spatial and velocity bias have the potential to systematically affect growth rate measurements from next generation surveys. In particular:
\begin{enumerate}
\item{Scale-dependent spatial bias has only a small effect on the growth rate constraints using the density field, as any systematic shift can be largely absorbed into the galaxy bias parameter. However, the addition of velocity field information means that this is no longer the case and the scale-dependent galaxy bias must be included. Fortunately, we find that doing so comes at little cost the growth rate constraints, whilst simultaneously providing interesting insight into the scale dependence itself.}\vspace{3pt}
\item{For next generation peculiar velocity surveys, velocity bias will be become an increasingly important effect. Neglecting it for the TAIPAN survey could bias growth rate constraints by over $5\sigma$ and marginalising over it can increase the error on $f\sigma_{8}$ by $\sim 40\%$. However, the effect of a velocity bias on the power spectrum, a reduction in small scale power, is similar to that provided by non-linear redshift space distortions, to the extent that any systematic biases can be compensated for by allowing suitable freedom in the redshift space model.}
\end{enumerate}}
\item{Finally, we see that for any reasonable magnitude of zero-point offset the growth rate constraints remain unbiased. This is because the effect of a zero-point offset is to add a constant shot-noise-like component to the power spectrum, which is also the case for the statistical observational errors in the peculiar velocity measurements. For any conceivable survey these observational errors far outweigh the systematic zero-point offset, such that this shot-noise term is statistically dominated and the growth rate measurements unbiased compared to the statistical uncertainty.}
\end{itemize}

Overall, this work has provided detailed forecasts for the cosmological information that can be inferred from joint peculiar velocity and redshift surveys. It has already been established that peculiar velocity measurements have the potential to improve constraints on the growth rate of structure obtained with redshift surveys, but we have shown that, in truth, peculiar velocity surveys have to potential to do more than simply augment density field analyses. For both the TAIPAN and WALLABY+WNSHS surveys, peculiar velocity samples will become a precise tool for cosmology in their own right. However, with such an increase in precision comes an increased requirement to understand the potential sources of systematic error.

\section*{Acknowledgements}
CH is grateful to Christina Magoulas and Matthew Colless for providing estimates of the number density of peculiar velocity targets for the TAIPAN survey, and Alan Duffy for making available his simulations related to the WALLABY+WNSHS survey. In addition to the aforementioned people, CH is also grateful to Chris Springob, Caitlin Adams, Andrew Johnson and Tamara Davis for their helpful insight and comments over the course of this research, and for the hospitality of the Centre for Astrophysics \& Supercomputing at Swinburne University of Technology. 

This research was conducted by the Australian Research Council Centre of Excellence for All-sky Astrophysics (CAASTRO), through project number CE110001020.

This research has made use of NASA's Astrophysics Data System Bibliographic Services and the \texttt{astro-ph} pre-print archive at \url{https://arxiv.org/}. All plots in this paper were made using the {\sc matplotlib} plotting library \citep{Hunter2007}

\end{document}